\documentclass{aa}

\usepackage{amssymb}
\usepackage{mathtools}
\usepackage{lscape}

\usepackage{natbib}

\usepackage{bm}

\usepackage{graphicx}
\usepackage{txfonts}
\usepackage{gensymb}
\usepackage{amsmath}
\usepackage{adjustbox}
\usepackage{multirow}
\usepackage{epstopdf}
\usepackage{tabulary}
\usepackage{tabularx}
\usepackage{hhline}
\usepackage{booktabs}
\usepackage{array}
\usepackage{subcaption}

\usepackage[all]{nowidow}
\usepackage{natbib,twoopt}
\usepackage[breaklinks=true]{hyperref} 
\bibpunct{(}{)}{;}{a}{}{,}             

\begin{document}

\title{Multi-Particle Collisions in Microgravity: Coefficient of Restitution and Sticking Threshold for Systems of Mm-Sized Particles}

\author{J. Brisset\inst{1}\inst{2}
          \and
          T. Miletich\inst{1}\inst{2}
          \and
          J. Metzger\inst{1}
          \and
          A. Rascon\inst{1}
          \and
          A. Dove\inst{1}
          \and
          J. Colwell\inst{1}
          }


\institute{Physics Department, University of Central Florida, 4000 Central Florida Boulevard, Orlando FL-32816
			\and
			Currently at: Florida Space Institute, University of Central Florida, 12354 Research Parkway, Orlando FL-32826
		   }


 
  \abstract
   {The current model of planet formation lacks a good understanding of the growth of dust particles inside the protoplanetary disk beyond mm sizes. A similar collisional regime exists in dense planetary rings. In order to investigate the low-velocity collisions between this type of particles, the NanoRocks experiment was flown on the International Space Station (ISS) between September 2014 and March 2016.  We present the results of this experiment.}
   {The objectives of our data analysis are the quantification of the damping of energy in systems of multiple particles in the 0.1 to 1 mm size range while they are in the bouncing regime, and the study of the formation of clusters through sticking collisions between particles.}
   {We developed statistical methods for the analysis of the large quantity of collision data collected by the experiment. We measured the average motion of particles, the moment of clustering, and the cluster size formed. In addition, we ran simple numerical simulations in order to validate our measurements.}
   {We computed the average coefficient of restitution (COR) of collisions and find values ranging from 0.55 for systems including a population of fine grains to 0.94 for systems of denser particles. We also measured the sticking threshold velocities and find values around 1 cm/s, consistent with the current dust collision models based on independently collected experimental data.}
   {Our findings have the following implications that can be useful for the simulation of particles in PPDs and planetary rings: (1) The average COR of collisions between same-sized free-floating particles at low speeds (< 2 cm/s) is not dependent on the collision velocity; (2) The simplified approach of using a constant COR value will accurately reproduce the average behavior of a particle system during collisional cooling; (3) At speeds below 5 mm/s, the influence of particle rotation becomes apparent on the collision behavior; (4) Current dust collision models predicting sticking thresholds are robust.}


\maketitle

\begin{table*}[tbp]
   \caption{List of the experiment cells in the NanoRocks tray and their content. *JSC-1 is a Lunar dust simulant \citep{mckay1994jsc}.}
   \centering
    \begin{tabular}{|c|c|c|c|c|}
    \hline
    \textbf{Tray} & \textbf{Tray Dimensions [mm]} & \textbf{Particle Diameter [mm]} & \textbf{Composition} & \textbf{Quantity} \\ \hline
    1     & 15x15x3 & 2     & Acrylic (red) & 15 particles \\ \hline
    \multirow{2}[4]{*}{2} & \multirow{2}[4]{*}{32x10x3} & 2     & Acrylic (red), breaded with chalk & 15 particles \\ \cline{3-5}
          &       & 0.87-1.18 & Glass (blue), breaded with chalk & 40 particles \\ \hline
    \multirow{2}[4]{*}{3} & \multirow{2}[4]{*}{32x10x3} & 2     & Acrylic (red) & 15 particles \\ \cline{3-5}
          &       & 0.87-1.18 & Glass (blue) & 40 particles \\ \hline
    \multirow{2}[4]{*}{4} & \multirow{2}[4]{*}{32x10x3} & 0.87-1.18 & Glass (blue), breaded with chalk & 40 particles \\ \cline{3-5}
          &       & 0.8-2 & Copper, breaded with chalk & 40 particles \\ \hline
    \multirow{2}[4]{*}{5} & \multirow{2}[4]{*}{32x10x3} & 0.87-1.18 & Glass (blue) & 40 particles \\ \cline{3-5}
          &       & 0.8-2 & Copper & 40 particles \\ \hline
    \multirow{2}[4]{*}{6} & \multirow{2}[4]{*}{15x15x3} & 0.87-1.18 & Glass (blue) & 30 particles \\ \cline{3-5}
          &       & 0.5-1 & JSC-1* & 0.4 g \\ \hline
    \multirow{2}[4]{*}{7} & \multirow{2}[4]{*}{32x15x3} & 2     & Acrylic (red) & 25 particles \\ \cline{3-5}
          &       & 0.5-1 & JSC-1* & 0.12 g \\ \hline
    8     & 15x15x3 & 0.5-1 & JSC-1* & 0.6 g \\ \hline
    \end{tabular}
   \label{t:trays}
 \end{table*}

\section{Introduction}
\label{s:intro}

The very first stages of planet formation involve the accretion and growth of dust particles inside protoplanetary disks (PPDs) orbiting young stars \citep{weidenschilling_cuzzi1993PP3, weidenschilling2000Icarus, dominik_et_al2007PPV}. Observation of very young PPDs ($<$ 1~million years) reveal the presence of $\mu$-sized grains that condensed as the disk is cooling \citep{Bouwman_et_al2007ApJ} while after about 1 Myr, the size of the grains has already reached a millimeter to a centimeter \citep{ricci2010dust,tazzari2016aap}. This observation of the grain growth inside PPDs is supported by numerical simulations \citep{okuzumi_et_al2009ApJ,wada_et_al2009ApJ} and experiments \citep{blum_wurm2008ARAA}, and the grain growth from $\mu$m- to mm-sized particles is currently well-understood. 

However, the processes of growth from the mm-sized particles to km-sized planetesimals (at which sizes gravity becomes the main acting force between individual bodies) remain unclear. Indeed, growth by sticking upon contact ceases for particles reaching millimeter sizes inside the PPD, as the collision regime transitions from sticking to bouncing and even fragmentation \citep{blum_wurm2008ARAA,guettler_et_al2010A&A,zsom_et_al2010AA}. One of the currently studied scenarios to overcome this "bouncing barrier" is the formation of rubble-pile planetesimals by the gravitational collapse of a highly concentrated particle cloud formed either inside turbulent vortices of the PPD gas or through streaming instabilities \citep{Johansen_et_al2007ApJ,johansen_et_al2014PP}. First models of such collapsing clouds currently oversimplify the collision behavior of the particles by, for example, assuming a coefficient of restitution of 0 for all particle collisions occurring during the collapse \citep{lorek2016comet}. Experimental studies of the collisional evolution of multi-particle systems composed of mm-sized particles are required to better understand the behavior of a collapsing particle cloud forming a planetesimal.

Experiments on the collision of sub-mm and mm-sized particles studied the transition between sticking and bouncing regimes, as well as aggregate fragmentation \citep{guettler_et_al2010A&A, weidling_et_al2012Icarus, kothe_et_al2013Icarus}. Most of these experiments detect individual collisions inside multi-particle systems observed under microgravity conditions and it is therefore difficult to collect a high amount of collision data. \citet{brisset_et_al2016_AA} introduced the statistical analysis of a high number of collisions between sub-mm sized SiO$_2$ aggregates recorded during the three minutes of microgravity provided by the REXUS suborbital rocket flight. This paper presents the data analysis and results of the NanoRocks experiment, which recorded a high number of collisions between mm-sized particles (one order of magnitude bigger in size) during the 18 months it remained in the microgravity environment of the International Space Station (ISS). The statistical analysis of these numerous collisions allows for the study of the dissipation of kinetic energy through bouncing collisions and the transition from bouncing to sticking regimes at these particle sizes.

The specific objective of the NanoRocks experiment was to study low-energy collisions of mm-sized particles of different shapes and materials. In particular, the experiment was designed to study the bouncing-to-sticking transition for collisions with decreasing collision velocity. From the dust collision model developed in \cite{guettler_et_al2010A&A}, this transition is expected at collision velocities around 1~cm/s for the particle sizes in NanoRocks. Such low collision velocities can only be reached in a microgravity environment.

The NanoRocks experiment was flown on the ISS between September 2014 and March 2016 as a NanoRacks payload. The experiment setup and performance are described in \cite{brisset2017nanorocks}. NanoRocks recorded collisions inside eight trays, each containing a different type of particles ranging from 0.1 to 1~mm in size (Table~\ref{t:trays}). Section~\ref{s:exp} briefly describes the NanoRocks experiment. Section~\ref{s:results} presents the data results obtained during the 18~months on-board ISS and Section~\ref{s:discussion} discusses these results and their applicability to planet formation and planetary ring particles.

\begin{figure}[tbp]
  \begin{center}
  \includegraphics[width = 0.48\textwidth]{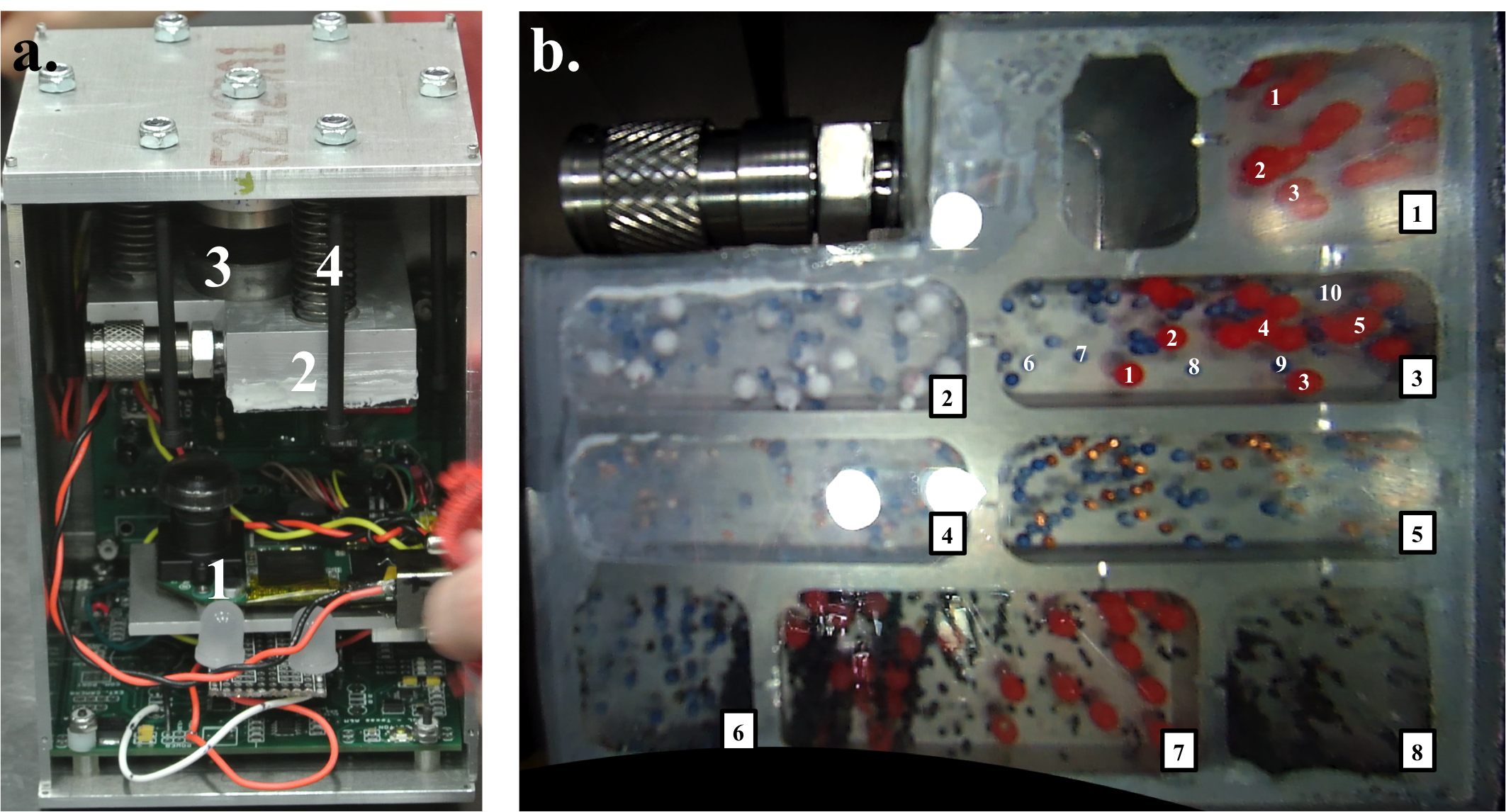}
 \caption{The NanoRocks experiment: a.) Experiment hardware before closing the flight unit: the camera and electronics can be seen at the bottom of the picture (1), while the tray (2) and its springs (4) and magnet (3) are at the top. b.) Recorded image of the experiment tray containing the eight particle samples during an experiment run on ISS. The experiment cell numbering is shown (black numbers in white boxes) as well as the particles tracked manually in each cell (white numbers).}
 \label{f:hw}
 \end{center}
\end{figure}

\section{The NanoRocks Experiment}
\label{s:exp}

The objective of the NanoRocks experiment is to study low-energy collisions of mm-sized particles of different shapes and densities. The main component of the experiment is an aluminum tray ($\sim$8$\times$8$\times$2~cm$^3$), which is divided into eight sample cells each holding different types and combinations of particles (Table~\ref{t:trays}). This tray is mounted on three springs to allow for its 3-dimensional shaking. Each 60~s, an electromagnet pulled the tray, compressing the springs and then releasing the tray leading to collisions between the particles and the tray walls. The inter-particle collisions generated by this shaking are recorded autonomously with a camera commanded by on-board electronics (Figure~\ref{f:hw}). Each experiment run consists of a 60-minute recording of the samples while they are being shaken once every minute. 

The different particles investigated (Table~\ref{t:trays}) were chosen to cover a range of densities (i.e. masses) and shapes, as well as surface texture, as described in \cite{brisset2017nanorocks}. The densities of acrylic, glass, and copper are 1.18, 2.60, and 8.96~g/cm$^3$, respectively. These values cover the range of densities of materials such as water ice, silicates, and metals, which are the main constituents of solid matter in PPDs and planetary rings. This variety of particles therefore allows for the study of the influence of mass on the sticking behavior, a parameter commonly used in dust collision models \citep[e.g.][]{guettler_et_al2010A&A}.

We targeted mm-sized particles in particular as it is the lower size range for which bouncing is expected to play an important role in PPDs, potentially stalling the growth of grains at these sizes \citep{zsom_et_al2010AA}. A better understanding of the bouncing behavior and transition to a sticking regime in this size range is essential for the accuracy of numerical simulations of dust growth in PPDs. In addition, we know from the study of primitive meteorites that mm-sized chondrules were present in the early Solar System \citep[possibly originating from the melting of dust aggregates, e.g.][]{hewins1997chondrules}. Chondrules played an essential role in the formation of early planetesimals and solid particles in the mm size ranges are reasonable analogs for these primitive particles.

Between September 2014 and March 2016, NanoRocks stayed on-board the ISS in the NanoRacks platform. During its time on orbit, the experiment performed nominally, recording low-velocity collisions inside the sample cells in 60-second video files of 96 MB each. Each video was recorded at a resolution of 1920$\times$ 1080 px and a speed of 30 fps. A total of 178 of these minute-long video files were downloaded from ISS and were used in the following data analysis. Details on the experiment hardware and performance are given in \citep{brisset2017nanorocks}.

\begin{table*}[t]
\begin{center}
\caption{Percentage of particles tracked automatically compared to the total number of particles in each particles tray. No JSC-1 particles could be tracked and tray 8 is therefore not included in the statistics. The percentage of tracked particles was measured for each shaking cycle and then averaged over all NanoRocks data downloaded from ISS. The mean absolute deviation between all the shaking cycles analyzed is also listed. Numbers over 100\% indicate that the tracking tool detected more particles than present (uneven spots on the images).}
\begin{tabular}{|c|r|r|r|r|r|r|r|}
\hline
Tray & \multicolumn{1}{c|}{1} & \multicolumn{1}{c|}{2} & \multicolumn{1}{c|}{3} & \multicolumn{1}{c|}{4} & \multicolumn{1}{c|}{5} & \multicolumn{1}{c|}{6} & \multicolumn{1}{c|}{7} \\ \hline
Percentage of tracked particles & 77.9574 & 110.743 & 104.618 & 24.1282 & 50.4868 & 49.2647 & 77.4346 \\ \hline
Mean deviation over all shaking cycles & 7.47332 & 12.1393 & 10.0224 & 7.983 & 4.20911 & 6.21834 & 6.07785 \\ \hline
\end{tabular}
\label{t:perc_auto_track}
\end{center}
\end{table*}

\begin{figure}[t]
  \begin{center}
  \includegraphics[trim=38 12 25 20, clip, width = 0.48\textwidth]{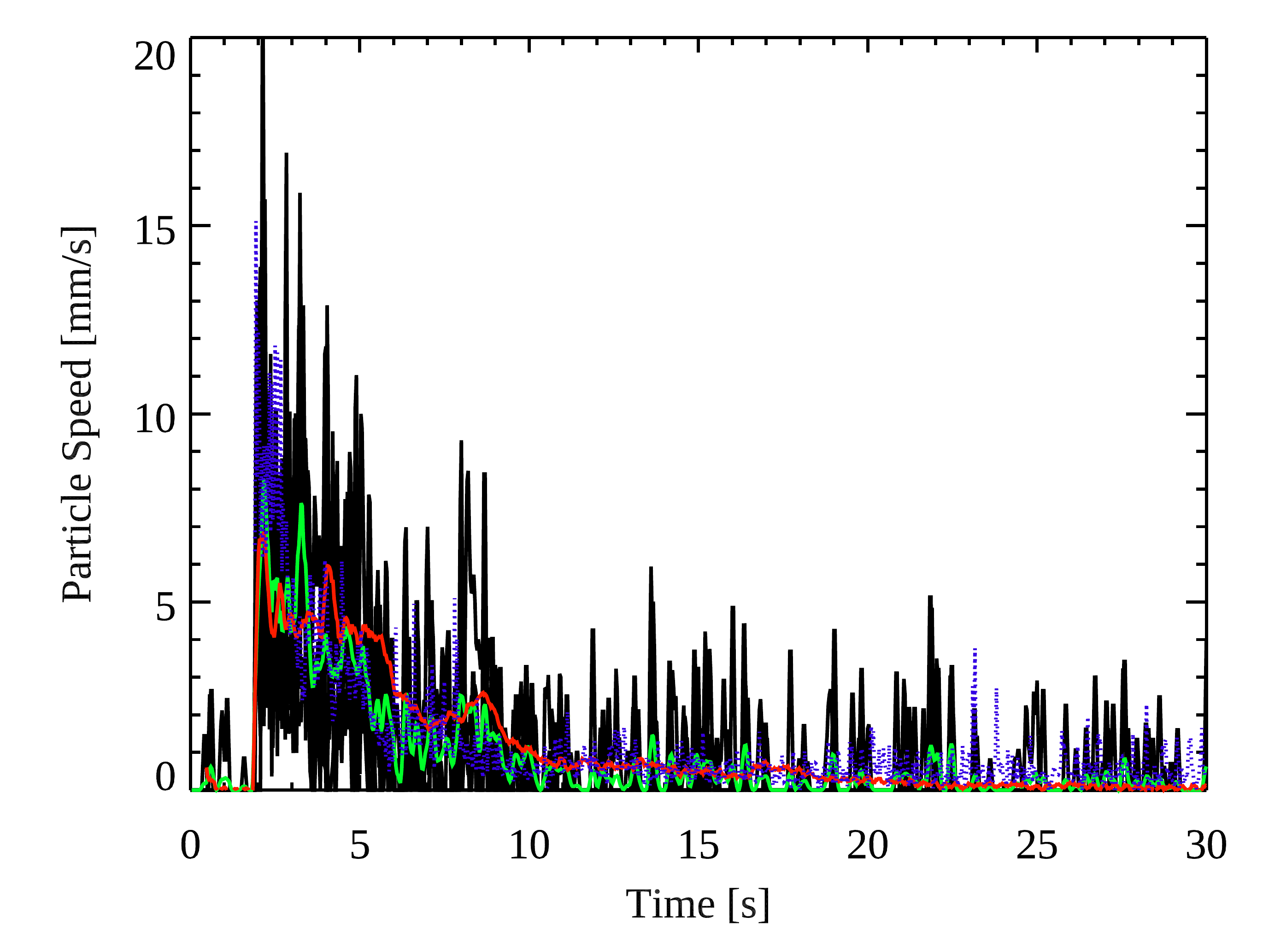}
 \caption{Particle speed measurement inside the NanoRocks tray 1 (Figure~\ref{f:hw}b and Table~\ref{t:trays}). The black curves show the speeds measured using manual tracking of 10 of the particles in tray 1. For each frame of the recorded video, the speed of these 10 particles was averaged and this average particle speed is shown by the green curve. The average particle speed obtained from the statistical measurement method after calibration (see text for details) is shown by the red curve. The average particle speed obtained from automated tracking is shown in blue (dotted).}
 \label{f:part_track}
 \end{center}
\end{figure}

\section{Scientific Data Results}
\label{s:results}

The video data retrieved from ISS shows the evolution of the different multi-particle systems in the NanoRocks trays in response to an input of kinetic energy in the form of a shaking event. During such a shaking event, the particles acquire speed through collisions with the tray walls. This acquired kinetic energy is then gradually lost through bouncing collisions with other particles and the experiment cell walls. When the average particle speed becomes low enough, inter-particle collisions lead to sticking of the collision partners, and particle clusters start forming. \\
The data analysis performed on these video files was performed along two lines of study:
\begin{itemize}
\item the kinetic energy damping through multiple collisions inside each of the NanoRocks particle trays. 
\item the cluster formation in each multi-particle system.
\end{itemize}
The following paragraphs develop the methods used for data analysis and the results obtained.

\subsection{Energy Damping through Multiple Collisions}

To study the damping of the kinetic energy inside each NanoRocks tray, the speed of the particles and its evolution over time after each shaking event had to be determined from the video data. Three different methods were applied to measure the particle speeds: manual particle tracking, automated particle tracking, and a statistical average particle speed determination. These methods and the speed measurement they delivered are described below.

\subsubsection{Analysis Methods}

The traditional way of determining particle speeds from video data is to track the position of individual particles, usually manually, during the relevant period of time \citep{weidling_et_al2012Icarus,kothe_et_al2013Icarus,brisset2017low}. This method is, however, not practical for the amount of data produced by NanoRocks and available for analysis. For this reason, two additional methods of determining the particle speeds during experiment runs were developed, and manual tracking was only used on a restricted set of data to validate and calibrate the measured speeds. These two methods consist of automated particle tracking and a statistical particle speed measurement. The following paragraphs describe all three analysis methods applied to the NanoRocks video data.

\paragraph{\textit{Manual Particle Tracking}}
Manual particle tracking was performed on a sample number of particles in the NanoRocks trays (Figure~\ref{f:part_track}) for one 60 s shaking cycle. Ten of each type of particle in each tray were tracked (e.g. ten red beads in tray 1, ten red beads and ten blue beads in tray 3, etc.), using the NASA Spotlight software \citep{klimek2006spotlight}.  The individual particle speeds during one experiment shaking cycle are derived from the tracked particle positions. An example is shown in Figure~\ref{f:part_track}, where the black curves are the speeds calculated from the manual tracks of 10 particles inside tray 1 (see Figure~\ref{f:hw}b), and the green curve is the average speed for these 10 particles. \\

\begin{figure}[t]
  \begin{center}
  \includegraphics[width = 0.48\textwidth]{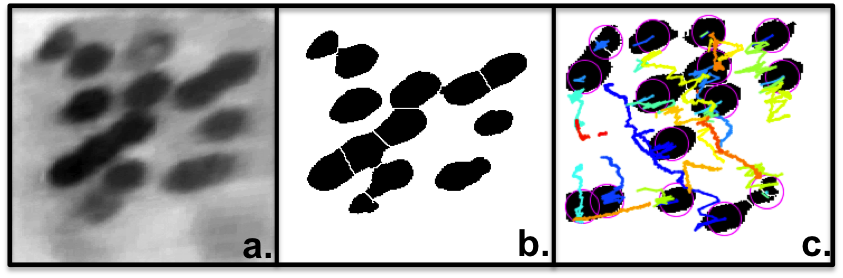}
 \caption{Automated particle tracking: (a) Processed image of tray~1 at the beginning of a shaking cycle. (b) The same image as in a.), binarized. (c) Binarized image of tray~1 at the end of a shaking cycle with the detected particles (purple circles) and tracks (colored lines) superposed. The track colors indicate the point in time during the shaking cycle: bluer colors indicate that the track was detected earlier during the shaking cycle while redder colors indicate later detections.}
 \label{f:auto_track}
 \end{center}
\end{figure}

\paragraph{\textit{Automated Particle Tracking}}
Ideally, each particle would be tracked manually and its speed and collision parameters measured directly. The high amount of data gathered by the experiment, however, requires more efficient analysis methods able to process all the available information. As a first step towards measuring all the particle position and speed information, the particles inside the NanoRocks trays were tracked in an automated fashion using the Trackmate plugin of the open source Fiji software \citep{metzger2016ASCE}. After a significant amount of image processing, this software was able to produce particle tracks for about 80\% of the particles that were not JSC-1. Figure~\ref{f:auto_track} shows the different steps leading up to the production of automated particle tracks in tray~1.

This tracking method greatly increased the number of particles that could be tracked in the NanoRocks data. However, the initial image processing steps were not entirely successful in allowing subsequent TrackMate detections, depending on the particle tray. For example, blue glass beads could not be detected in trays 2, 3 and 4 due to their low contrast compared to the aluminum tray background, and JSC-1 particles were not detected at all (trays 6, 7, 8). Table~\ref{t:perc_auto_track} lists the percentage of tracked particles compared to the total number of particles inside each NanoRocks tray. For most trays, only an incomplete set of particles could be detected by the tracking tool. For trays 2 and 3, the image processing led to the tracking tool detecting background brightness unevenesses as particles, resulting in percentages higher than 100\%. We concluded that automated tracking was reliable only in trays 1 and 5.

Because of this lack of precision and in order to gain an better and more complete data set from the NanoRocks images, we developed an additional speed detection method based on a statistical analysis of the recorded images.

\begin{figure}[t]
  \begin{center}
  \includegraphics[trim=6 6 4 11, clip, width = 0.45\textwidth]{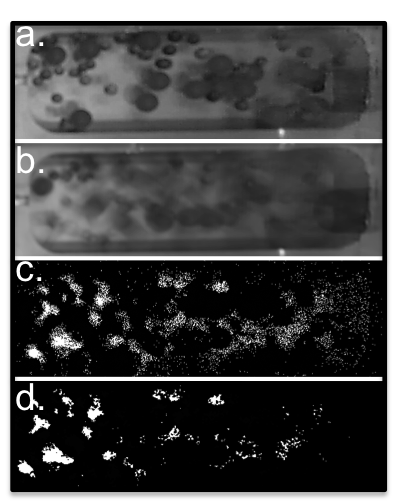}
 \caption{Processing of the NanoRocks data frames for statistical data analysis: (a) the gray frame consists of a mixture of the three (red, green and blue) layers of the original frame, chosen to distinctly reveal the red and blue beads. (b) This averaged frame is produced by averaging 10 consecutive gray frames. (c) The next step consists in subtracting each frame from its following frame. (d) Finally, the subtracted frame gets despeckled and binarized.}
 \label{f:proc}
 \end{center}
\end{figure}

\paragraph{\textit{Statistical Data Analysis}}
\label{s:stat}

The statistical method to measure the average particle speed in the NanoRocks trays is based on the detection of particle motion between subsequent data images. In a first step, the color frames recorded by the NanoRocks camera need to be processed in order to detect particle motion in the experiment cells. After correcting the fish-eye distortion of the lens, each data frame is converted from color to 8-bit grayscale (Figure~\ref{f:proc}a). In order to make movement of the particles apparent, each frame is then averaged over 10 frames (Figure~\ref{f:proc}b). Particle motion is now visible in the form of blurred out trajectories. To follow the evolution of particle motion during the experiment run, two consecutive averaged frames are then subtracted one from another (Figure~\ref{f:proc}c). In a final step, residual noise is reduced by despeckling and binarizing the subtracted frames (Figure~\ref{f:proc}d).

\begin{figure*}[t]
  \begin{center}
  \includegraphics[width = \textwidth]{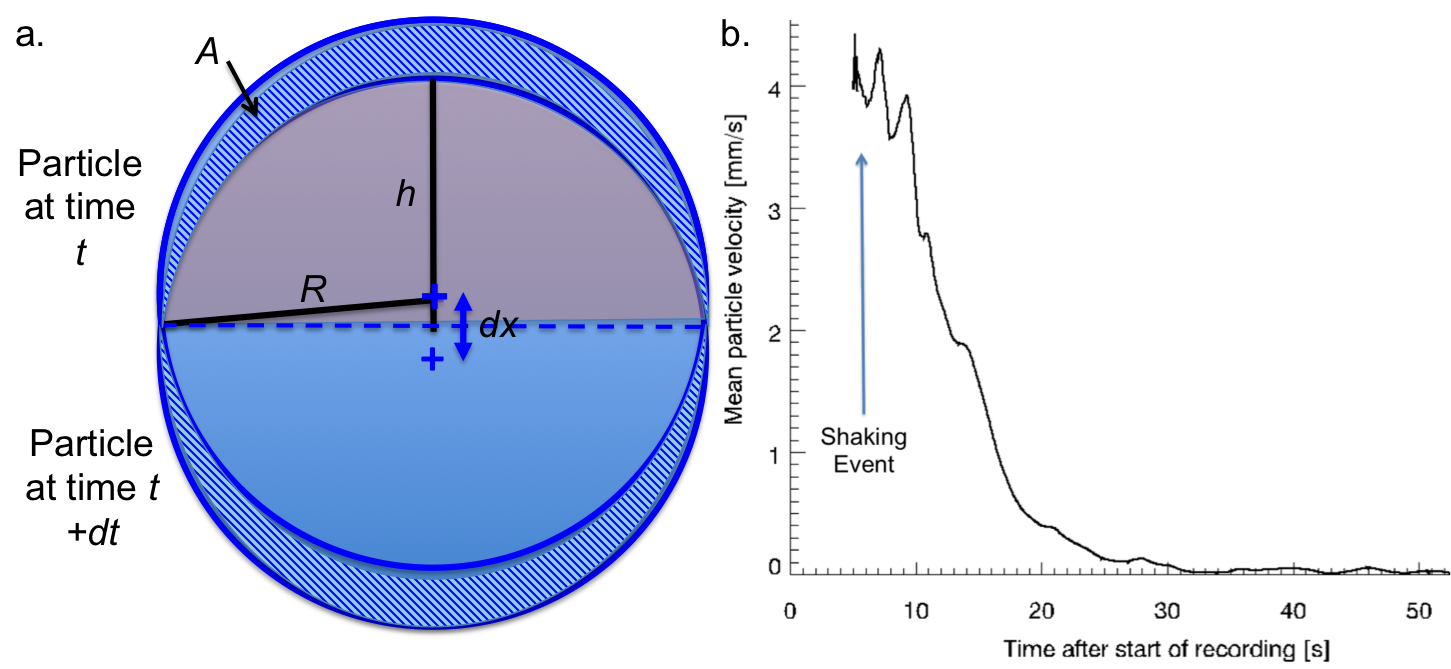}
 \caption{Statistical measurement of the average particle velocity inside the NanoRocks experiment cells during a shaking cycle: a) Definition of the quantities used to link the particle moving surface to the average particle speed: between the times $t$ and $t+dt$, the particle center displacement is $dx$. The particle is spherical with a radius $R$ and the height of the red circle portion is defined as $h$. The area highlighted by the processing described in Section~\ref{s:stat} is striped on this schematic. b) Average particle speed inside the NanoRocks tray 3 during one shaking cycle, measured as described in the text.}
 \label{f:stat_vel}
 \end{center}
\end{figure*}

The highlighted surface in the images processed in that way now marks and quantifies particle motion. During a shaking event, particles are moving very fast, leading to a large highlighted surface. After damping of the particle velocities following a shaking event, the particles are barely moving, leading to a very small highlighted surface. \\
This measured highlighted surface can be linked to the average particle speed by considering that it corresponds to the difference in surface covered by particles from one frame to another (striped area in Figure \ref{f:stat_vel}a, for one particle). Basic geometrical considerations result in this surface being (for one particle)
\begin{equation}
A=2\pi R^2-4R^2\left[\arccos{(1-\frac{h}{R})}-(1-\frac{h}{R})\sqrt{\frac{h}{R}}\sqrt{2-\frac{h}{R}}\right]
\label{e:geometry}
\end{equation}
If we write $1-\frac{h}{R}=\frac{dx}{2R}$ and consider that $\frac{dx}{2R}\ll 1$, we can perform a Taylor expansion on Equation~\ref{e:geometry} resulting in $A=4Rdx$. As $dx=vdt$, $v$ being the average particle speed, and for a number $N$ of particles, we have
\begin{equation}
v=\frac{NA}{4NRdt}
\label{e:av_speed}
\end{equation}
$NA$ is the measured surface in the last step of the image processing (Figure~\ref{f:proc}d.). The average particle velocity in a tray can be then computed by inserting the corresponding values for $N$, $R$ (see Table~\ref{t:trays}) and $dt=0.033$~s (30 fps video recording). Figure~\ref{f:stat_vel}b. shows an example of the average velocity in the experiment cell~3 (see Figure~\ref{f:hw}b.) during one shaking cycle. The shaking event happens about 5~s after the start of the video recording and the particle mean velocity gets damped over the next 20~s by inter-particle collisions. As shown in the following paragraph, this method allowed us to capture the evolution of particle motion in a more reliable manner than the automated tracking method.

\subsubsection{Results}

\paragraph{\textit{Average Particle Speed during an Experiment Run}}
\label{s:speed}

Figure~\ref{f:part_track} shows an example of the statistically computed average particle speed inside tray~1 during one of the shaking cycles recorded (red curve). We found that the statistical speed measurement method values had a systematic offset compared to manual tracks. This offset was due to the detection of motion of tray background features and side walls, which are not included in Equation~\ref{e:av_speed}. As these particle-unrelated motion detections are random, it is difficult to consider them analytically. Instead, a simple calibration using the manually measured average velocities allows us to correct for them. After this calibration, the statistically measured velocity profiles match well with the average speeds computed via manual and automated tracking. This result was valid for all NanoRocks trays except for trays containing chalk dust where the agglomeration of dust on the tray walls and corners added the tray motion to the particle motion detected. This was corrected with a simple calibration of the statistically computed speed using the manual tracks. Figure~\ref{f:part_track} also shows the recorded speeds using the automated tracking method (blue curve). In tray 1, most particles were detected so that these results can be compared to our other speed determination methods. We find that automated tracking and manual tracking are in good agreement. The combination of all three methods supports the validity of our statistical velocity determination method after calibration.

\begin{figure}[t]
  \begin{center}
  \includegraphics[trim=38 12 25 20, clip, width = 0.48\textwidth]{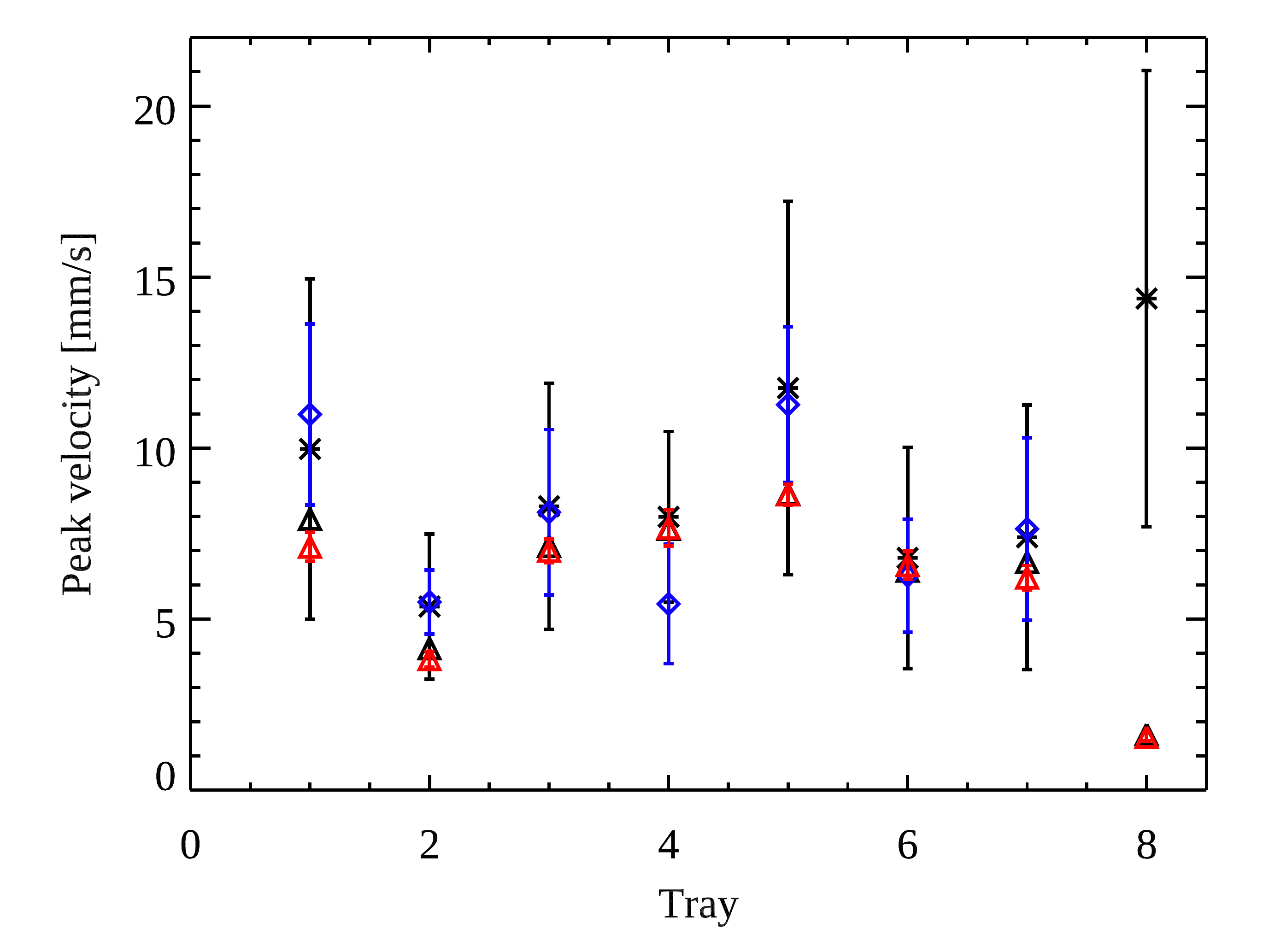}
 \caption{Peak velocity inside the NanoRocks trays: measured manually for one shaking cycle (averaged over 10 particles, black asterisks); computed via the automated tracking (averaged over all tracks, blue diamonds); and with the statistical method described in the text (triangles). The velocity values determined statistically are shown for the data set that was also tracked manually (black triangles), and averaged over 178 shaking cycles (red triangles with the mean absolute deviation shown as error bars (red). Red and black triangles overlap for most trays.}
 \label{f:max_vel}
 \end{center}
\end{figure}

During each shaking cycle, the particle velocity peaks at the moment of the shaking event (at around 7~mm~s$^{-1}$ in Figure~\ref{f:part_track}). The average velocity of particles then decreases over the next $\sim$20~s due to the energy loss induced by inter-particle collisions. For the rest of the shaking cycle, particles barely move and no further collisions take place.

Figure~\ref{f:max_vel} shows the peak velocity of the particles inside the NanoRocks trays averaged over all the data retrieved. This peak velocity is reached at the shaking event (see Figure~\ref{f:part_track}). Asterisks show the manually measured peak velocities, and diamonds show the ones averaged over all automated tracks. Triangles show peak velocities calculated using the statistical method. Manual and automated tracking values are in good agreement except for tray~4. The low percentage of particles detected in this tray (see Table~\ref{t:perc_auto_track}) can most likely account for this discrepancy. We note that the statistical method systematically underestimates the velocity values. In order to verify that this was not due to the number of experiment runs considered to perform the measurement, Figure~\ref{f:part_track} shows the statistical velocities both for all data sets (red triangles) and only the one that was tracked manually (black triangles). As the red and black triangles overlap within their error bars for all trays, we can conclude that the experiment runs reliably generated the same particles speeds across a recording time of several months. The systematic offset between the manual tracking and statistical values was found to be due to a measurement bias in the manual tracking. As motion is easiest detected for the fastest moving particles, these were preferentially tracked during manual tracking. The statistical analysis on the other hand records all particle motion, so that the average of the recorded speeds includes slower particles compared to the manual tracking. This effect is most acute for the tray with only JSC-1 particles (tray 8), where most particles were tied up in clumps that were too strong to be de-aggregated during the shaking events. The clumps themselves were moving at slow speed, while the particles remaining outside of clumps (the ones that were manually tracked) had higher speeds. As automated tracking relied on the detection of moving particles, it also preferentially tracked faster particles, displaying a good match to the manual data.

Particles in trays where chalk dust was present reached lower peak velocities than their counterparts without chalk: the tray 2 maximum speeds are lower than for tray 3 (acrylic and glass beads), and tray 4 particles reached lower speeds than in tray 5 (glass and copper beads). As the particles receive their kinetic energy from the moving cell walls during a shaking event, we suspect that these lower peak velocities were due to the fine chalk particles coating the aluminum walls. Such coating reduces the coefficient of restitution between the incoming particle and the wall, and part of the wall's kinetic energy is lost in the collision instead of being entirely transferred to the particle.

\begin{figure}[t]
  \begin{center}
  \includegraphics[trim=16 20 20 45, clip, angle=90, width = 0.48\textwidth]{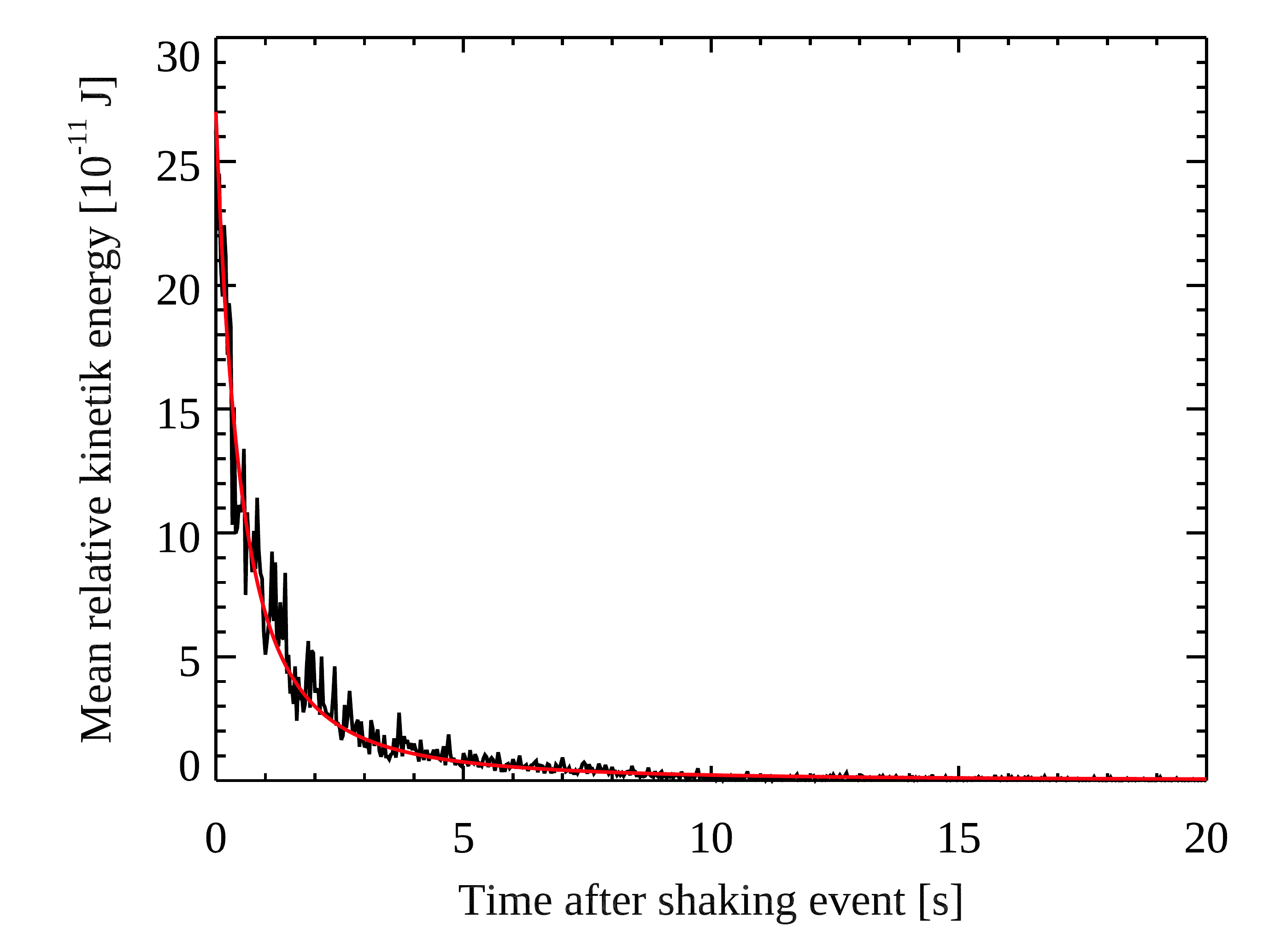}
\begin{picture}(0,0)
\put(-200,50){\includegraphics[height=5.0cm]{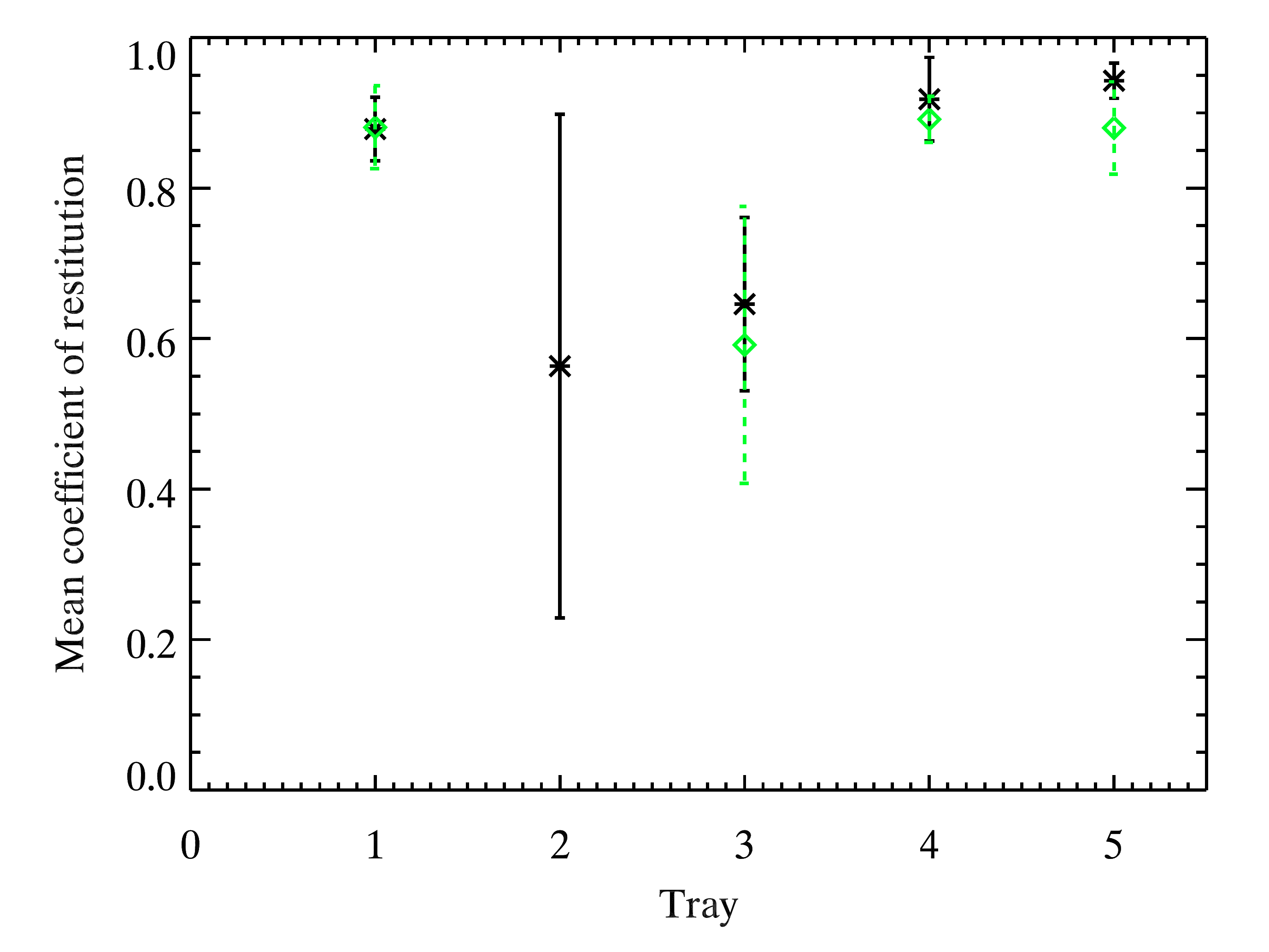}}
\end{picture}
 \caption{Mean relative kinetic energy in the NanoRocks tray 1 after a shaking event. The red curve shows the $T(t) = \frac{T_0}{(1+\frac{t}{\tau_0})^2}$ fit with the parameters $T_0=2.70\times10^{-10}$~J and $\tau_0=1.64$ s. The computed mean coefficient of restitution for this shaking cycle is $\epsilon=0.89$. The inset shows the mean coefficient of restitution for each NanoRocks tray calculated using these fit parameters averaged over 178 shaking cycles (black asterisks, see Table~\ref{t:trays} for a tray description). When possible, we also calculated the average coefficient of restitution using the same method based on the automated tracking data (green diamonds).}
 \label{f:kin_en}
 \end{center}
\end{figure}

\paragraph{\textit{Determination of the Average Coefficient of Restitution}}
\label{s:cor}

The decay in average particle speed measured above is due to multiple collisions between particles in each tray. Therefore, the measurement of the average particle speed can be linked to the average loss of kinetic energy and hence to the coefficient of restitution of each collision.

Previous work by \citet{heisselmann2010} has demonstrated that a multi-particle system in bouncing collision mode behaves the same if the distribution of the coefficient of restitution for each collision is uniform around a mean value $\epsilon_0$ or if each collision happens at the same and constant coefficient of restitution $\epsilon_0$. From their data (1.5~cm-sized particles between 6 and 22 cm/s) and the bouncing collision data collected by \citet{weidling_et_al2012Icarus} (mm-sized particles between 1 and 10 cm/s), we conclude that it is reasonable to assume a uniform coefficient of restitution for bouncing collisions between mm-sized particles at relative velocities below about 1 cm/s. In a first simplification, we will use a constant coefficient of restitution for the particles inside the NanoRocks trays during a shaking cycle, the value of which will represent the average of such a uniform distribution.

\begin{table*}[th]
\begin{centering}
\caption{Results of the manual analysis of clusters formed inside the NanoRocks trays. The values were averaged over 18 shaking cycles and used to calibrate the statistical analysis. Particles were counted regardless of their composition (see Table~\ref{t:trays} for a description of the particles in the different trays).}
\begin{tabular}{|c|>{\centering\arraybackslash}p{1in}|>{\centering\arraybackslash}p{1.5in}|>{\centering\arraybackslash}p{1.75in}|>{\centering\arraybackslash}p{1.5in}|}
\hline
\multicolumn{ 1}{|c|}{Tray} & \multicolumn{ 4}{c|}{Manual cluster analysis} \\ \cline{ 2- 5}
\multicolumn{ 1}{|c|}{} & Average number of clusters & Average number of particles per cluster & Mean absolute deviation for the number of particles per cluster & Percentage of particles involved in clusters \\ \hline
1 & 3.44 & 2.85 & 0.73 & 65.4 \\ \hline
2 & 6.78 & 7.20 & 2.77 & 88.7 \\ \hline
3 & 8.44 & 5.52 & 1.36 & 84.8 \\ \hline
4 & 6.67 & 7.70 & 2.09 & 85.6 \\ \hline
5 & 9.61 & 6.55 & 1.34 & 78.7 \\ \hline
6 & 5.78 & 97.54 & 21.95 & 93.9 \\ \hline
7 & 6.56 & 42.17 & 9.14 & 98.7 \\ \hline
8 & 4.83 & 133.01 & 44.96 & 80.4 \\ \hline
\end{tabular}
\label{t:cluster}
\end{centering}
\end{table*}

In order to extract the mean coefficient of restitution of the collisions in a tray, we will use Haff's law for cooling granular gases \citep{haff1983JFM}: \\
\begin{equation}
	T(t) = \frac{T_0}{(1+\frac{t}{\tau_0})^2}
\label{e:temp_def}
\end{equation}
where $t$ is the time and $T$ is the temperature of the collisional multi-particle system given by the mean kinetic energy per particle:
\begin{equation}
	T=\frac{1}{2}m<v>^2
\label{e:temp}
\end{equation}
$m$ being the average mass and $<v>$ the average velocity of a particle. $T_0$ is the initial temperature of the multi-particle system and $\tau_0$ the characteristic temperature decay time. According to Haff's law \citep{haff1983JFM}, the parameter $\tau_0$ depends on the mean coefficient of restitution $\epsilon$ as follows \citep[e.g. ][]{maass2008experimental}:
\begin{equation}
	\frac{1}{\tau_0}=\sigma n (1-\epsilon^2) <v_0>
\label{e:tau}
\end{equation}
where $\sigma=\pi (2R)^2$ is the collision cross-section of a particle in the tray ($R$ is the radius of a particle), $n=\frac{N}{V}$ is the particle number density ($N$ is the number of particles in the tray and $V$ the volume of the tray) and $<v_0>$ the initial mean velocity of the particles. By measuring the evolution of the mean particle velocity in NanoRocks trays over time, we can calculate the average coefficient of restitution of collisions inside this tray.

To this purpose, we computed the mean relative kinetic energy inside the NanoRocks trays using Equation~\ref{e:temp}, the mean particle velocity and the particle characteristics of each tray listed in Table~\ref{t:trays}. As the velocity required by Equation~\ref{e:temp} is the mean relative velocity between colliding particles, we used work performed in \citet{brisset_et_al2016_AA} that calculated the probabilities of collisions happening at relative velocities ranging from 0 to $2v_{mean}$ in a tray containing particles moving at a mean velocity $v_{mean}$. This work shows that, for each collision, the relative velocity between particles has the highest probability to be $2v_{mean}$. As we are using statistical particle velocity data to compute an average coefficient of restitution, it is appropriate to use this highest probability value for the relative velocity of particles colliding in the NanoRocks trays.

Figure~\ref{f:kin_en} shows the computed mean relative kinetic energy of the particles in tray~1 (see Table~\ref{t:trays}). A function of the type $T(t)=\frac{T_0}{(1+\frac{t}{\tau_0})^2}$ was fitted to this relative kinetic energy and the mean coefficient of restitution was extracted via
\begin{equation}
	\epsilon=\sqrt{1-\frac{1}{\sigma n}\sqrt{\frac{m}{2}}\frac{1/\tau_0}{\sqrt{T_0}}}.
\end{equation}
In the case of Figure~\ref{f:kin_en}, $m=4.9$~mg, $\sigma=\pi(2R)^2=1.26\times10^{-5}$~m$^2$ and $n=\frac{N}{V}=2.22\times10^7$~m$^{-3}$ are the particle mass, collision crosssection and number density, respectively. These numbers yield an average coefficient of restitution of 0.89.\\
The computed coefficients of restitution for each shaking cycle recorded were averaged giving a mean coefficient of restitution for each NanoRocks tray shown in the inset of Figure~\ref{f:kin_en}.

Trays 2 and 4 contain the same particles as trays 3 and 5 respectively, with the addition of chalk dust. The presence of dust in the multi-particle system reduces the mean coefficient of restitution of collisions. This is expected, as particles coated with dust have shown to have lower coefficients of restitution than smooth-surfaced particles \citep{beitz_et_al2012Icarus}.

\subsection{Cluster Formation and Morphology}
\label{s:morph}

\subsubsection{Manual Cluster Analysis}

For 18 of the shaking cycles of the NanoRocks payload (the first 18 videos received from ISS), we manually analyzed the particle clusters that formed inside each tray after the kinetic energy of the particles induced by the shaking event was dissipated by collisions. This analysis consisted in counting the number of clusters formed and the number of particles each cluster was composed of. Table~\ref{t:cluster} lists the average number of clusters and number of particles per cluster measured. Particles were counted regardless of their composition (acrylic, glass, copper or JSC-1). For the purpose of this analysis, we considered any sticking agglomerate of two or more particles to be a cluster. JSC-1 formed big clusters in the corners of the tray. As the particle color, size, and number made it impossible to distinguish each particle, the number of JSC-1 particles in these clusters was estimated based on the size of the cluster and the known total number of particles present in the tray. For trays containing JSC-1 particles, we estimated the number of JSC-1 particles from the mass and known density and porosity of grains.

Due to the high amount of NanoRocks data available, this manual counting of clusters and particles was not applied to every shaking cycle recorded. We therefore developed a statistical method to measure the average number of particles in clusters at the end of each shaking cycle.

\subsubsection{Statistical Cluster Analysis}

In addition to the manual counting of a sample number of formed clusters, we performed a statistical analysis of these clusters over all the NanoRocks data downlinked from ISS. This analysis described below delivered the average size of clusters (number of composing particles) formed in each particle tray at the end of each shaking cycle.

\begin{figure}[t]
  \begin{center}
   \begin{subfigure}[b]{1\textwidth}
   \includegraphics[width=0.5\textwidth]{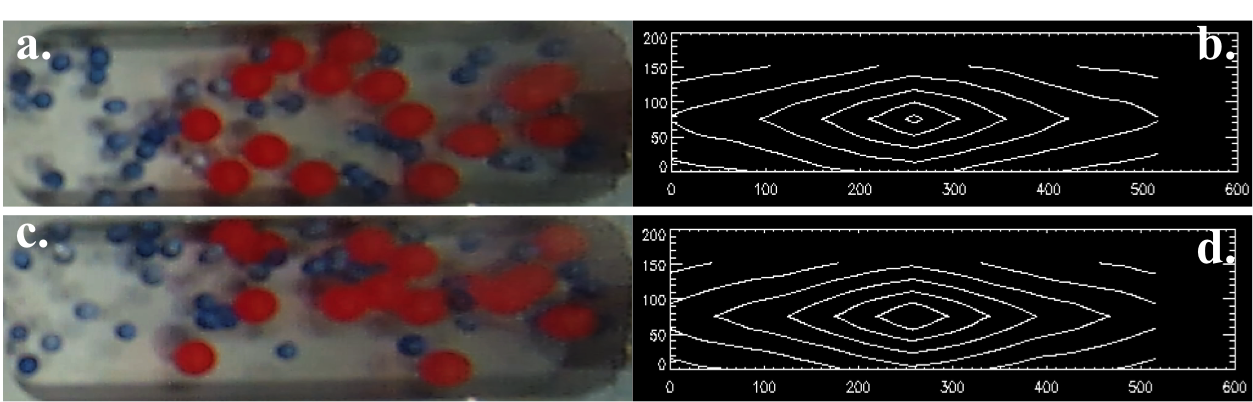}
\end{subfigure}
\begin{subfigure}[b]{1\textwidth}
   \includegraphics[trim=38 25 25 8, clip, width=0.48\textwidth]{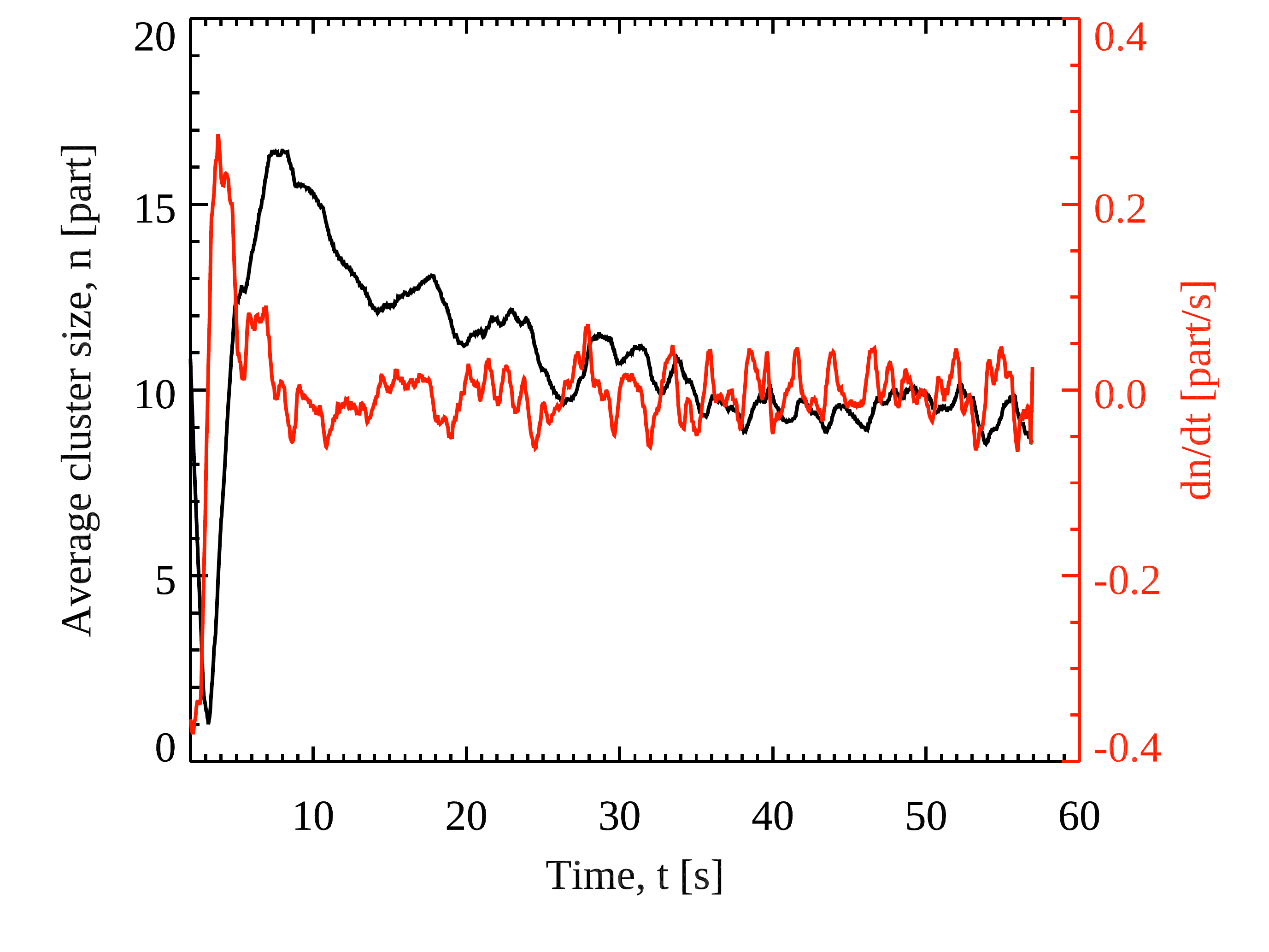}
\end{subfigure}

 \caption{Example of the image autocorrelation performed on the NanoRocks data. During the shaking event, the particles in tray 3 (a) are colliding individually, bouncing off each other, while 20 s after the shaking event, very slow collisions between particles have led to the formation of clusters (c). Comparing the autocorrelation results of these two images, contoured using the same (arbitrary) levels (b and d, respectively), gives a quantitative information on the final cluster size. (e) Calibrated cluster size deduced from the measurement of the area enclosed in an arbitrary autocorrelation level during a NanoRocks shaking cycle (see text for details) in a recording of tray 3 containing 2-mm red acrylic and 1-mm blue glass beads (see Table \ref{t:trays}). In red we show the derivative of the cluster size with time dn/dt, which can be used to detect the onset of clustering in the tray.}
 \label{f:autocorr}
 \end{center}
\end{figure}

A common technique to determine the particle size scales of clusters in N-body simulations is the use of an autocorrelation function \citep[e.g.,][]{daisaka1999spatial}. This function quantifies the similarities between an image and itself when swiped along its two dimensions. The highest level of correlation is therefore always located at the center of the image. If particles are individually distinct from each other in this image, the autocorrelation will be very sharp at the center. However, if particles form clusters, the autocorrelation, while still maximal at the center, will be broader and reflect the shape and size of the clusters present. Figure~\ref{f:autocorr} shows an example of image autocorrelation for particles in the NanoRocks tray 3 (see Table~\ref{t:trays}). The top tray image (a) was captured during the shaking event when particles are agitated and bounce off each other in collisions. The contour plot of the autocorrelation of this image (b) shows that the highest levels of autocorrelation are reached at the center of the image (the contour levels are arbitrary). The bottom tray image (c) was captured 20 s after a shaking event, when the particles' energy was dissipated and very low-energy collisions led to particle sticking and cluster formation. The contour plot of the autocorrelation of this second image (d), using the same level values, shows that the presence of clusters has increased the autocorrelation values around the center. The presence and size of clusters can be detected by tracking autocorrelation levels in images over a complete NanoRocks shaking cycle.

To detect the formation and presence of clusters, an arbitrary autocorrelation level is chosen and the number of pixels (area) for which the autocorrelation value is higher than this level are counted for each image of a NanoRocks shaking cycle. The shape of the curve obtained can be seen in Figure~\ref{f:autocorr}e for one experiment run in tray~3 (the shaking event happens at 3~s in the chosen recording): at the time of the shaking event, the autocorrelation values of the images are at their minimum, as only individual particles are present in the tray. During the 6~s following the shaking event, the autocorrelation values of the recorded images increase, as the particle kinetic energy decreases and more and more collisions result in sticking, hence forming first particle binaries, then aggregates and clusters. At about 10~s, the autocorrelation values reach a maximum as clusters in the tray have formed. The arbitrary pixel area corresponding to this observed maximum represents the mean area of a cluster in the tray. To obtain the quantitative size of these clusters via the autocorrelation method, we use the manual cluster analysis as calibration (already applied in Figure~\ref{f:autocorr}e). In the case of Figure~\ref{f:autocorr}e, the maximum cluster size is 16.5 particles per cluster.

\begin{figure}[t]
  \begin{center}
  \includegraphics[trim=20 10 20 38, clip, width = 0.36\textwidth , angle=90]{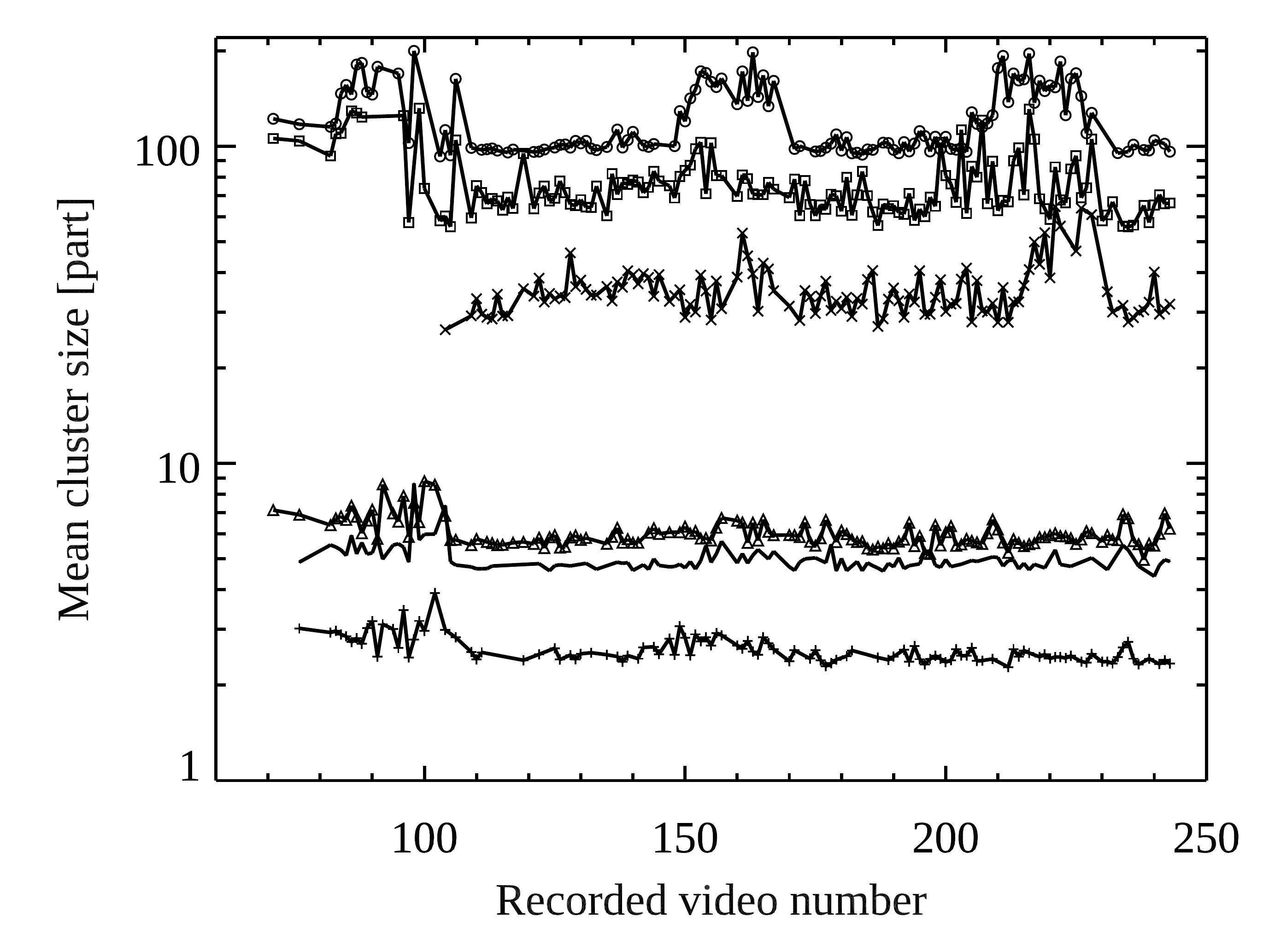}
 \caption{Long-term evolution of the average cluster size in each tray over the lifetime of the experiment. Each data point shows the cluster size after a shaking event and subsequent energy damping through multiple collisions leading to clustering. Dates span from November 2014 (left) to July 2015 (right). The video labels as shown (about 70 to 250) do not correspond to regular time intervals. No offset is used for the different trays. From bottom to top, we show tray 1 (plus signs), 3 (no symbols), 5 (triangles), 7 (x), 6 (squares), and 8 (circles).}
 \label{f:cluster_evol}
 \end{center}
\end{figure}

\subsubsection{Results}

\paragraph{\textit{Average Cluster Size}}

The results obtained by the manual cluster analysis can be found in Table~\ref{t:cluster}. For the manual analysis we were able to count both the number of particles in each cluster and the number of clusters formed. This allowed to estimate the number of particles involved in clusters (average number of clusters $\times$ average number of particles per cluster) and compare it to the total number of particles present in each tray (see Table~\ref{t:trays}). This computed percentage of particles involved in clusters is also listed in Table~\ref{t:cluster}.

In the manual analysis, we can see that trays 2 and~4 formed clusters with a higher number of particles than trays 3 and~5. This is expected as trays 2 and~4 contained the same particles as trays 3 and~5 but coated with chalk dust (see Table~\ref{t:trays}). The presence of the chalk dust increases the stickiness of the particles leading to larger clusters. The fact that the number of clusters counted for tray~4 is lower than in tray~5 is due to the reflection of the NanoRocks LEDs covering part of the tray during experiment runs and reducing the number of visible particles and clusters (see Figure~\ref{f:hw}). For this reason, we adjusted the total number of particles we used to compute the percentage of particles involved in clusters to the "visible" number of particles, using the surface percentage covered by the LED reflection to estimate the percentage of particles hidden by this reflection.

In Figure~\ref{f:cluster_evol}, we show the long-term evolution of the average cluster size measured using the autocorrelation method. Cluster sizes are consistent with the manual counting as expected after calibration. We can see that there is no general trend of increasing cluster sizes over the months spanned by the experiment runs. However, we can identify three periods in which clusters grew larger in all the trays. No precise time logs are available to determine the exact times these recordings were performed (NanoRocks was not keeping an absolute time and recording as soon as powered on). It is possible that these recordings were performed during quieter times on-board ISS, perhaps during sleep periods. As the NanoRocks tray was accommodated on springs, it was very sensitive to any type of vibration and activity levels. Sleep time experiment runs would have resulted in a lower ambient acceleration environment and larger final cluster sizes.

In trays~2 and~4, the autocorrelation cluster analysis was not able to produce meaningful results. This was due to the presence of white chalk in these trays that generated a poor contrast between the particles and the aluminum background, so that the autocorrelation of images could not efficiently detect shapes inside the tray. 

\begin{figure}[t]
  \begin{center}
  \includegraphics[trim=30 13 10 20, clip, width = 0.5\textwidth]{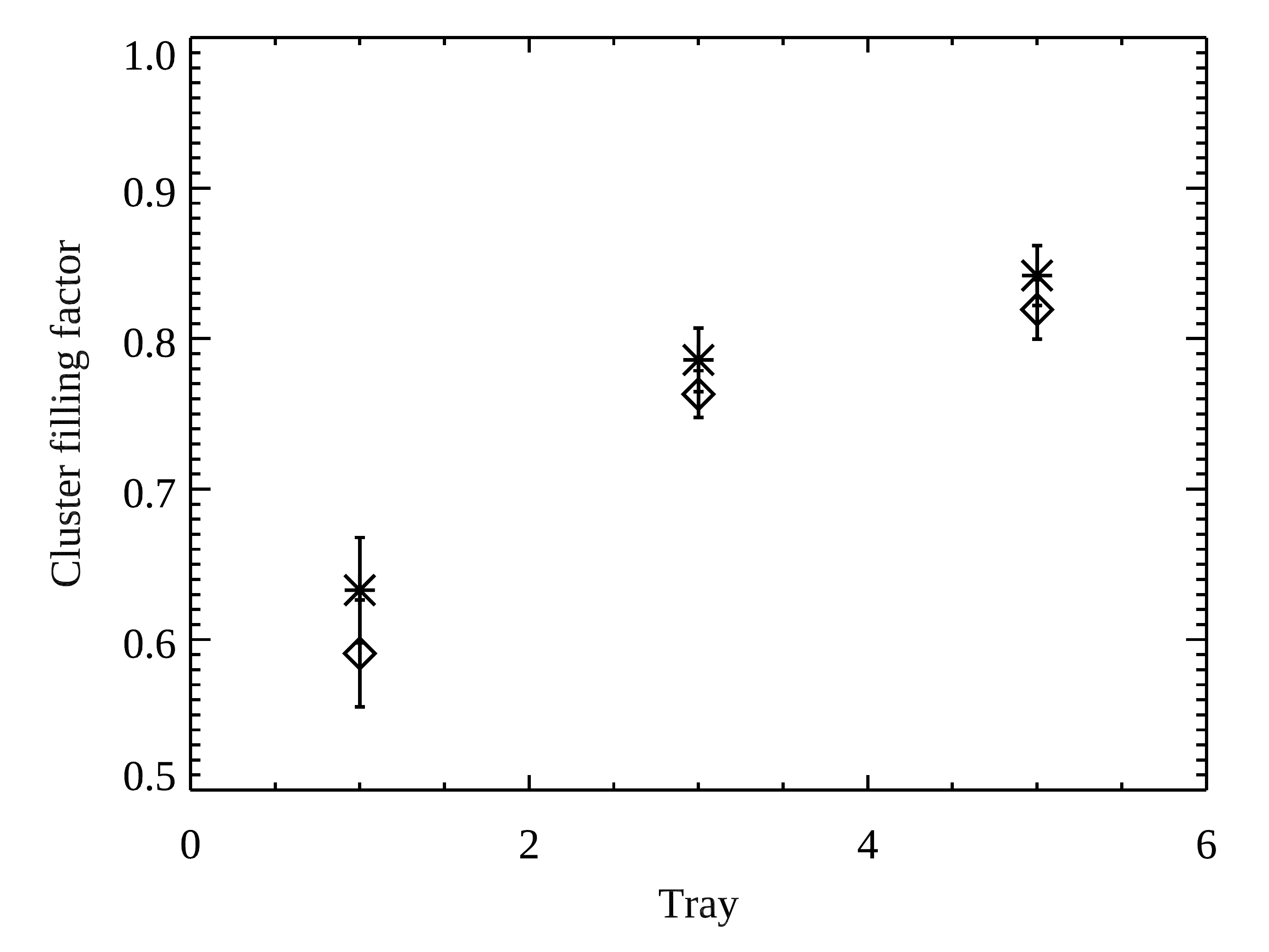}
 \caption{Average filling factor of clusters in trays 1 (spherical acrylic beads, 2 mm in diameter), 3 (spherical acrylic and glass beads, 2 and 1 mm in diameter, respectively), and 5 (spherical glass and ovoid copper beads, both about 1 mm in diameter). Diamonds show the minimum cluster filling factor immediately after clustering onset, and asterisks show the final filling factor after compaction through wall collisions.}
 \label{f:ff}
 \end{center}
\end{figure}

\paragraph{\textit{Evolution of the Cluster Shape}}

In Figure~\ref{f:autocorr}e we can see that, after this first maximum, the average cluster size decreases again before settling at a final value (10 particles in this example). This feature was observed in most of the other experiment runs and trays. Given the low relative speeds of the particles after clusters have started forming, this decrease cannot be due to the loss of particles through cluster fragmentation which would require much more energetic collisions. Instead, this apparent decrease in the cluster size is due to the compaction of the clusters through collisions with the cell walls. After each shaking event, the springs of the shaking mechanism display residual motion for 20 to 30~s, as motion damping due to the weight of the tray is not effective in microgravity. The apparent size of compacted clusters is smaller than for fractal structures, so that our autocorrelation measurement method detects a gradual cluster compaction as a decrease in size.

In order to quantify this compaction, we estimated the cluster filling factor from the autocorrelation data. For this, we assumed that the clusters grown in the NanoRocks trays were quasi-two-dimensional, thus only one particle thick ($d$ height, $d$ being the particle radius). We used the calibrated average area of clusters ($A$) measured through autocorrelation to calculate the average cluster volume, $V_{clus} = Ar$. From the manual cluster analysis we know the average number of particles in clusters ($nb_p$) and can therefore calculate the volume occupied by particles in the average clusters: $V_{part} = nb_p \frac{m_p}{\rho_p}$, with $m_p$ the mass of a particle and $\rho_p$ its density. $m_p$ and $\rho_p$ are known particle properties \citep[see ][]{brisset2017nanorocks}. For trays with several types of particles, the average mass and density were used. By dividing both volume quantities we obtain the estimated filling factor for the average clusters, $ff = V_{part}/V_{clus}$. If we consider that the number of particles in clusters remains the same during compaction, the decrease in detected cluster size results in an increase of the filling factor. Figure~\ref{f:ff} shows the results of this estimation, in trays 1, 3, and 5, for which autocorrelation results were reliable and particle numbers were actually counted rather than estimated (as for trays 6, 7, and 8 containing JSC-1 grains).

We note that the final filling factor for the spherical acrylic beads is around the Random Close Packing (RCP) value of 64~\%, which indicates that our filling factor estimation is of the right order of magnitude. If we assume that the pressure from the wall collisions is about the same in each tray, differences in the estimated filling factors are due to the nature of the clusters and their constitutive particles. We can see that clusters composed of particles with two different sizes (tray~3) are more compact than for same-sized ones (tray~1), as well as clusters containing non-spherical particles (tray~5). The values observed in trays 3 and 5 are consistent with the highest packing densities for binary sphere systems \citep{de2014upper,hopkins2012densest}.

In Figure~\ref{f:ff}, diamonds indicate the minimum cluster filling factor immediately after clustering onset, and asterisks show the final filling factor after compaction through wall collisions: in average, the filling factors increased from the diamond to the asterisks values after the onset of clustering in each experiment run. The compactions observed ($\Delta ff/ff_{min}$) are 6.6, 2.9, and 2.7 \% in trays 1, 3, and 5, respectively.

\paragraph{\textit{Particle Speed at Onset of Clustering}}
\label{s:onset}

In addition to the cluster sizes and shapes, the autocorrelation profile in the NanoRocks trays during shaking cycles allows for the detection of the time of particle clustering onset. The profile shown in Figure~\ref{f:autocorr}e displays a minimum value that corresponds to the shaking event, followed by rapid cluster growth upon the onset of clustering. In order to detect this onset, we used the derivative of the calibrated cluster size, shown in red in Figure~\ref{f:autocorr}e. The moment this derivative becomes positive was chosen as onset of clustering. In order to determine the average speed of the particles at that moment, we used the statistically computed particle speed (see Section~\ref{s:speed}). This allowed for the determination of an average relative speed of the particles (twice the average particle speed) at the moment of clustering onset. We obtain values of 10.4, 10.9 and 12.8~mm/s in trays 1, 3 and~5, respectively. 

\section{Discussion}
\label{s:discussion}

Through the data analysis described above, we monitored the evolution of the multi-particle systems inside the NanoRocks trays, and from this data we extracted the mean coefficient of restitution between particles and the threshold relative velocity of colliding particles at the onset of clustering. These two parameters determine the evolution of the particle system after each shaking event. In order to verify our data analysis approach, we performed computational simulations reproducing the NanoRocks multi-particle systems and compared our experimental results to the simulated systems produced. This allowed us to evaluate one of the key assumptions of our analysis: the use of a constant coefficient of restitution for collisions between particles during homogeneous cooling phases. In the following we discuss our results and their implications for system evolution in planetary environments.


\subsection {Validating our Data Analysis Approach}

In order to additionally validate our statistical data analysis approach, we designed and ran computer simulations of the multi-particle systems of NanoRocks. The simulation code was written in C\# and recreated the motion of a set of particles during homogeneous cooling through inter-particle collisions. In this custom code, we used the simplest approach of a hard-sphere model and an event-driven calculation of the particle positions and speeds. For each tray and shaking sequence, we determined three types of input parameters:
\begin{itemize}
\item Hardware and sample defined properties: the number of particles inside the tray, the particle radius, the tray dimensions and a range of initial speeds for particles (strength of the shaking event). The particle and tray properties were defined by the experiment hardware and the samples introduced before sealing the trays. The initial particle speeds were chosen to have a uniform distribution with the mean value measured for the peak average velocities in the NanoRocks data (see Figure~\ref{f:max_vel}).
\item Random parameters: initial position and velocity direction of the particles.
\item Variable parameters: average coefficient of restitution for inter-particle collisions and velocity threshold for sticking between particles (onset of clustering).
\end{itemize}

Collisions with the tray walls were considered to be perfectly elastic. Collisions between particles can have only two outcomes: either the particles bounce off each other with a coefficient of restitution set to the same value for all the collisions, or the particles stick together and have a new velocity determined by the conservation of momentum. The simulation picks the collision outcome by comparing the relative velocity of the two colliding particles to a threshold sticking velocity.

\paragraph{\textit{Verifying Measured Velocity Profiles}}

Due to the random nature of the particle positions and velocity directions at the start of the homogeneous cooling in both the simulations and the experiments, the direct comparison between simulation runs and experimental results is of limited value. Instead, we compared results averaged over 178 simulation and experiment runs (178 experiment runs were available in the NanoRocks data set). Therefore, for every tray and set of variable parameters considered, we ran 178 simulations and compared the average particle velocity profile to the average profile measured using our experimental data. The simulations were three-dimensional and the measured velocity profiles were derived from the two-dimensional projections of the particles and their motion onto the plane perpendicular to the line of sight, thus mimicking our experimental measurements using 2D video data.

\begin{figure}[t]
  \begin{center}
  \includegraphics[width = 0.5\textwidth]{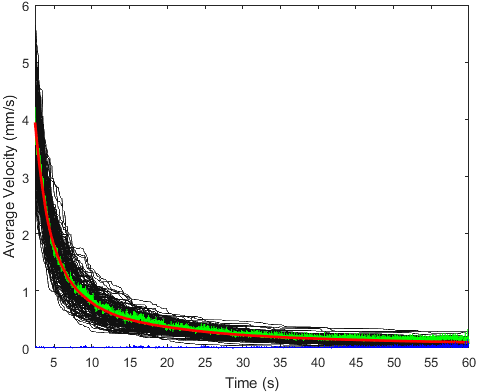}

  \begin{picture}(0,0)
  \put(-50,100){\includegraphics[height=3.9cm]{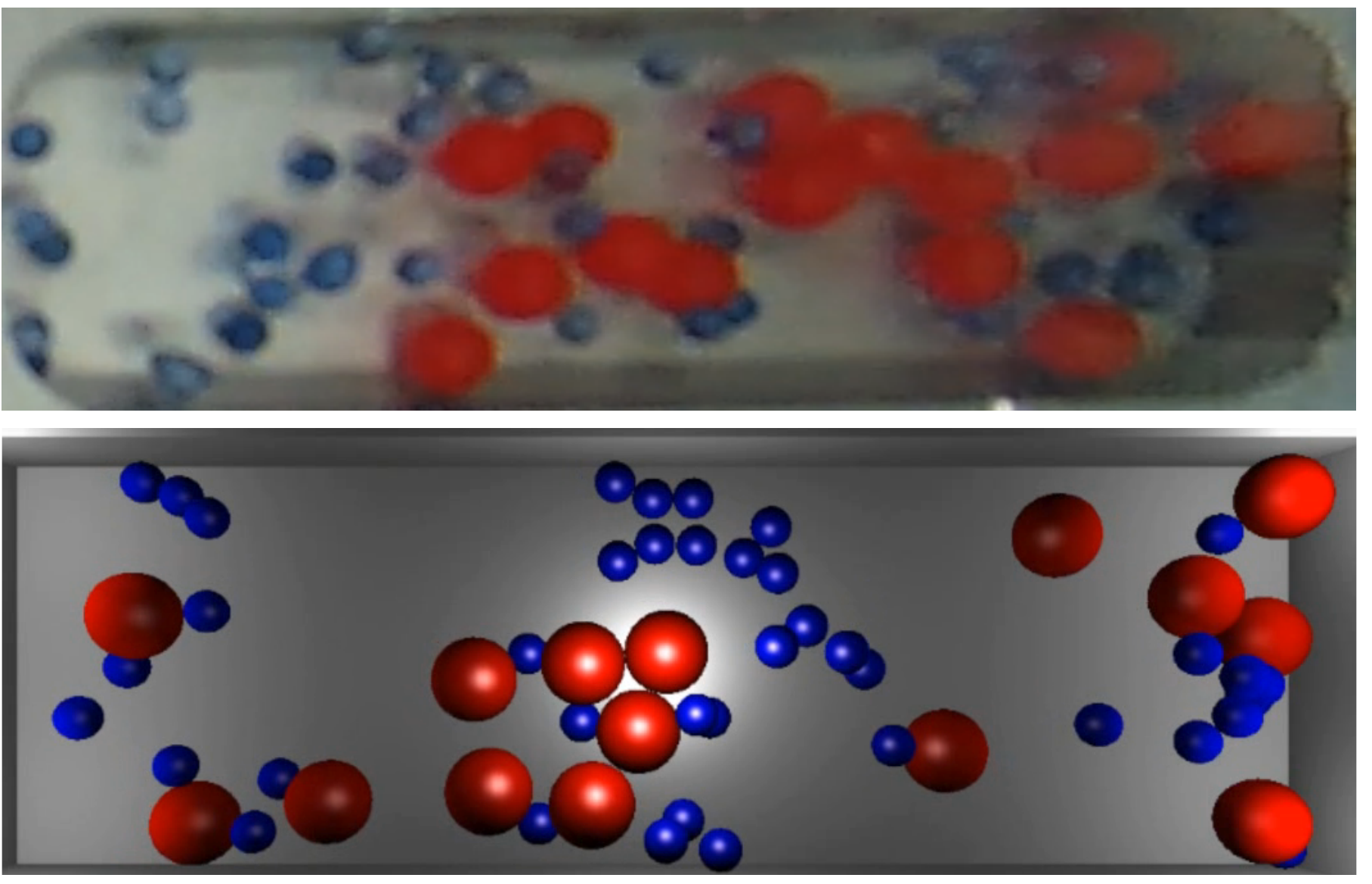}}
  \end{picture}
  
 \caption{Computational simulation results for the particle velocity evolution over time inside tray~1. Black curves show individual simulation runs and the red curve the average over 178 runs (the number of experimental runs available in the NanoRocks data set). The green curve shows the average  particle speed inside tray~1 computed with the statistical analysis (see Section~\ref{s:speed}). The shaking event was shifted to match the time of simulation start. The $\chi^2$ deviation between the simulated and measured average curves is shown in blue (mean value $7.10^{-3}$). The coefficient restitution for collisions between particles was set to 0.89 and for collisions with the cell walls at 0.99. The inset shows a visual of the simulation in tray~3 (bottom) compared to the NanoRocks data (top).}
\label{f:sim_graph}
 \end{center}
\end{figure}

\begin{figure*}[t]
  \begin{center}
  \includegraphics[width = \textwidth]{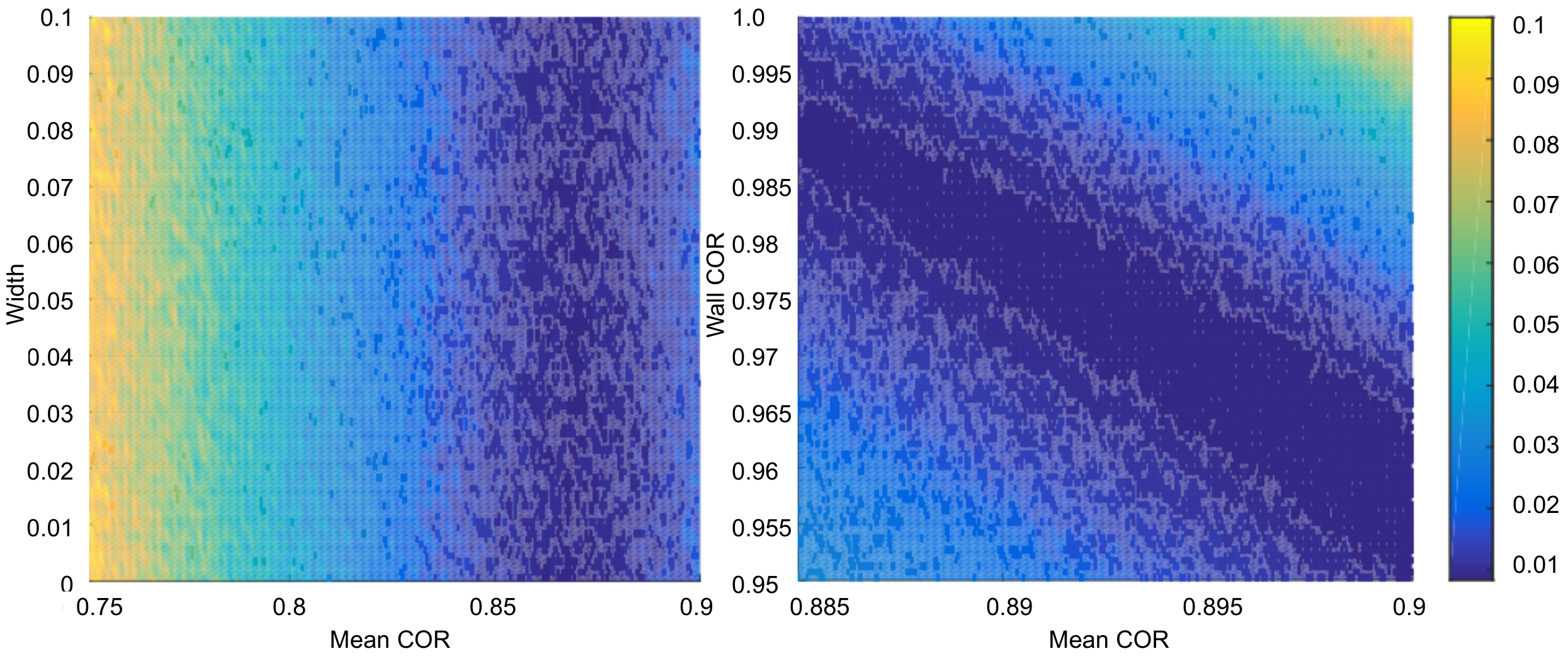}
 \caption{Examples of the search for optimal simulation parameters to match the NanoRocks experiment data. The simulation parameters varied were the average coefficient of restitution and width of the uniform distribution for inter-particle collisions and the coefficient of restitution for particle-wall collisions (considered constant). The colors indicate the average $\chi^2$ between the average simulation and experiment data (red and green curves in Figure~\ref{f:sim_graph}, respectively), with bluer colors indicating better matches. We search the parameter field for the average inter-particle coefficient of restitution (COR) and the particle-wall coefficient of restitution (Wall COR, right) and between the COR and the width of the COR uniform distribution (left). We observe a degeneracy in the optimal matches with several parameter configurations leading to the same velocity profile as observed in the NanoRocks experiment.}
 \label{f:param_search}
 \end{center}
\end{figure*}

Figure~\ref{f:sim_graph} shows an example of simulation results for tray~1. The velocity profiles in the 178 simulations are shown in black and their average in red. The average velocity profile measured in NanoRocks is shown in green, and the $\chi^2$ deviation between the simulated and measured average curves is shown in blue. The shown simulation result is one of the best $\chi^2$ match for our experimental data in the parameter ranges explored. They were obtained using a constant inter-particle coefficients of restitution of 0.89. The wall collisions were considered to have a coefficient of restitution of 0.99. This matching set of parameters was found by running a search over a range of parameters. Figure~\ref{f:param_search} shows two examples of parameter field searches for matches to the NanoRocks data. The left panel shows variations in the choice of a uniform distribution for the coefficient of restitution between particles (COR) with the distribution mean and width as parameters. The right panel shows variations in the choice of coefficients of restitution for the inter-particle collisions and the collisions between particles and the tray walls (Wall COR). Dark blue parameter combinations show the better matches to the measured NanoRocks data. We can see that the best matches are not unique and several combinations of parameters can reproduce the particle velocity evolution observed in the experimental data. 

The right panel of Figure~\ref{f:param_search} shows that the best matches are not obtained with elastic wall collisions, which was our original assumption in the statistical data analysis. We therefore revised equations \ref{e:temp_def}, \ref{e:temp}, and \ref{e:tau} to include the kinetic energy loss induced by collisions with the cell walls. Equation \ref{e:temp_def} is derived from the integration of the variation of kinetic energy at each time step, $\frac{\Delta T(t)}{\Delta t}$. In our assumption of elastic wall collisions, we assumed that the loss of kinetic energy was due solely to inter-particle collisions, so that the energy loss rate is calculated using

\begin{equation}
	\Delta T_p(t) = \frac{1}{2}m(1-\epsilon_p^2)<v>^2 ; \Delta t_p = \frac{1}{n\sigma <v>}
\end{equation}

with $v$, the average velocity of particles. $\Delta t_p$ is the average time between two inter-particle collisions. $\epsilon_p$ is the coefficient of restitution of inter-particle collisions (named COR in Figure~\ref{f:param_search}). In the same way, we can define the loss of kinetic energy due to collisions of particles with the cell walls and the corresponding collision time:

\begin{equation}
	\Delta T_w(t) = \frac{1}{2}m(1-\epsilon_w^2)<v>^2 ; \Delta t_w = \frac{2L}{<v>}
\end{equation}

with $L$ the side of the cell. $\Delta t_w$ is the average time between two particle-wall collisions. $\epsilon_w$ is the coefficient of restitution of particle-wall collisions (named Wall COR in Figure~\ref{f:param_search}). We obtain the total energy loss rate by adding particle-particle and wall-particle contributions:

\begin{multline}
	\frac{\Delta T(t)}{\Delta t} = \frac{\Delta T_p(t)}{\Delta t_p} + \frac{\Delta T_w(t)}{\Delta t_w} \\
	= m\left[n\sigma(1-\epsilon_p^2) + \frac{1}{4L}(1-\epsilon_w^2)\right]<v>^3
\end{multline}

If we integrate this equation as performed to derive equations \ref{e:temp_def} and \ref{e:tau}, we obtain the same form for the kinetic energy $T(t)$ as in equation \ref{e:temp_def}, with a new term in $\frac{1}{\tau_0}$ quantifying the influence of particle-wall collisions:

\begin{equation}
	\frac{1}{\tau_0} = \left[\sigma n(1-\epsilon_p^2) + \frac{1}{2L}(1-\epsilon_w^2)\right]<v_0>
\end{equation}

with $<v_0>$ the initial relative velocity between particles. If we solve for the coefficients of restitution as performed in \ref{s:cor}, we obtain

\begin{equation}
	\begin{split}
		& \left(\frac{\epsilon_w}{a}\right)^2 + \left(\frac{\epsilon_p}{b}\right)^2 = 1 \\
		& a = b\sqrt{2L\sigma n}\\
		& b = \sqrt{1+\frac{1}{2L\sigma n}-\frac{1}{\sigma n}\sqrt{\frac{m}{2}}\frac{1/\tau_0}{\sqrt{T_0}}} \\
	\end{split}
\label{e:ellipse}
\end{equation}

We can recognize an ellipse of semi-major axis $a$ and semi-minor axis $b$. Using the same numerical values as in \ref{s:cor}, we calculate $a = 0.95$ and $b = 2.8$. The values for the coefficients of restitution obtained by our numerical simulations, namely $\epsilon_p = 0.89$ (COR) and $\epsilon_w = 0.99$ (Wall COR) are in good agreement with equation~\ref{e:ellipse}. In the right panel of Figure~\ref{f:param_search}, we can see a portion of the ellipse defined by equation~\ref{e:ellipse}.

In conclusion, we were able to reproduce the NanoRocks experiment in its homogeneous cooling phase in numerical simulations. Our simulation results confirmed that the particle-wall collisions were quasi-elastic ($\epsilon_w = 0.99$), so that our first assumption of elastic wall collisions is validated. The evolution of the average particle velocity in the experiment was well reproduced by the simulations, thus validating our statistical velocity measurement approach.

\paragraph{\textit{The Constant Coefficient of Restitution Assumption}}

During our statistical data analysis (\ref{s:cor}), we assumed a constant coefficient of restitution over the range of collision velocities occurring during the homogeneous cooling of the NanoRocks particle system. Our computational simulations allowed us to evaluate the sensitivity of our results to this assumption. 

As mentioned in \ref{s:cor}, previous experiment results indicate that, at low collision velocities ($\leq 1$ cm/s), coefficients of restitution between particles of mm to cm in size display a uniform distribution around an average that is constant with the collision velocity. As a simplifying approach, we chose to perform our statistical data analysis using a constant coefficient of restitution. In our numerical simulations, we were able to determine the influence of a uniform distribution of coefficients of restitution on the resulting particle velocity evolution. We varied both the average inter-particle coefficient of restitution and the width of the distribution. On the left panel of Figure~\ref{f:param_search}, we can see that the velocity profile matches between simulated and experimental data do not depend on the width of the distribution of coefficients of restitution. This means in particular that the assumption of a constant coefficient of restitution is equivalent to the assumption of a uniform distribution with the same average, thus validating our simpler approach in~\ref{s:cor}.

The collision velocity dependence of the coefficient of restitution between particles was observed in experiments \citep{bridges1996energy,higa1996measurements} and is discussed in the literature \citep[e.g., ][]{ramirez1999coefficient,zhang2002modeling}. However, the experiments by \cite{heisselmann2010microgravity}, used in the present paper to support our assumption of a constant coefficient of restitution, do not see a variation of the coefficient of restitution between particles at low collision velocities ($\leq$ 1 cm/s). This discrepancy in results might originate in the nature of the collisions studied in these different experiments: \cite{bridges1996energy} and \cite{higa1996measurements} performed collisions of a particle with a flat surface, while \cite{heisselmann2010microgravity} observed particle-particle collisions in a free-floating environment. The latter is an experimental environment very similar to NanoRocks. In such inter-particle collisions in free-floating environments, other physical effects lead to a different behavior of the energy dissipation during collisions. In particular, the damping behavior of a large plate or surface is expected to differ from that of a same-sized particle, so that the velocity dependence of the coefficient of restitution might be an effect of the experimental setup in \cite{bridges1996energy} and \cite{higa1996measurements}. \cite{colwell2016low} and \cite{brisset2018regolith} studied collisions between a round cm-sized particle and a flat surface of fine grains. They also observed an increase of the coefficient of restitution with decreasing collision velocity. While the composition of the target surface was different than in \cite{bridges1996energy} and \cite{higa1996measurements} (fine granular material vs. solid ice), the similar behavior of the coefficient of restitution supports the fact that particle-surface collisions are very different from particle-particle collisions, and coefficients of restitution are only velocity dependent for collisions with particles with very different sizes (a much larger particle can be approximated as a target surface).

In addition, particle rotation while free-floating can lead to transfers between rotational and translational kinetic energy that will be reflected in the coefficient of restitution of a collision, leading to a distribution of values. In particular at very low relative velocities, microscopic surface asperities can play a significant role in the redistribution of kinetic energy between these two components, thus increasing the width of the distribution of values for the coefficient of restitution. In a first approximation (\ref{s:cor}), we used a constant coefficient of restitution for our data analysis of the NanoRocks collisions. In order to investigate a potential evolution of the coefficient of restitution with the collision speed, we revisited our experimental data. As a starting point, we used the average velocity profiles measured in our statistical analysis (\ref{s:cor}), which were validated by our simulations as described above. From the instantaneous velocity in each tray and knowing the particle number and size, we can deduce the current collision frequency at each time of our data collection (each frame of the video recording): $F_{c,k} = n\sigma v_k$, with $n$ and $\sigma$ as defined above, and $v_k$ the average particle velocity computed by our statistical analysis at the video frame $k$. Given that the frame rate at which the NanoRocks data was recorded is 30 fps, the average number of collisions experienced by particles between two recorded frames ($k$ and $k+1$) is $\frac{F_{c,k}}{30}$. We note that for the typical particle speeds in NanoRocks, the collision rate ranges from about 0.3 to 2 collisions per second. Therefore, we averaged the below measurements over about 30 frames in order to include at least one collision in the time of measurement. If $\epsilon_k$ is the coefficient of restitution for these collisions, which can be considered constant during one time step (no more than one collision is occurring between each frame), we have $\frac{v_{k+1}}{v_k} = \epsilon_k^{\frac{F_{c,k}}{30}}$, obtaining

\begin{equation}
	\epsilon_k = e^{\frac{30}{F_{c,k}}ln\frac{v_{k+1}}{v_k}}
\label{e:inst_cor}
\end{equation}

\begin{figure}[t]
  \begin{center}
  \includegraphics[trim=30 15 20 10, clip, width = 0.48\textwidth]{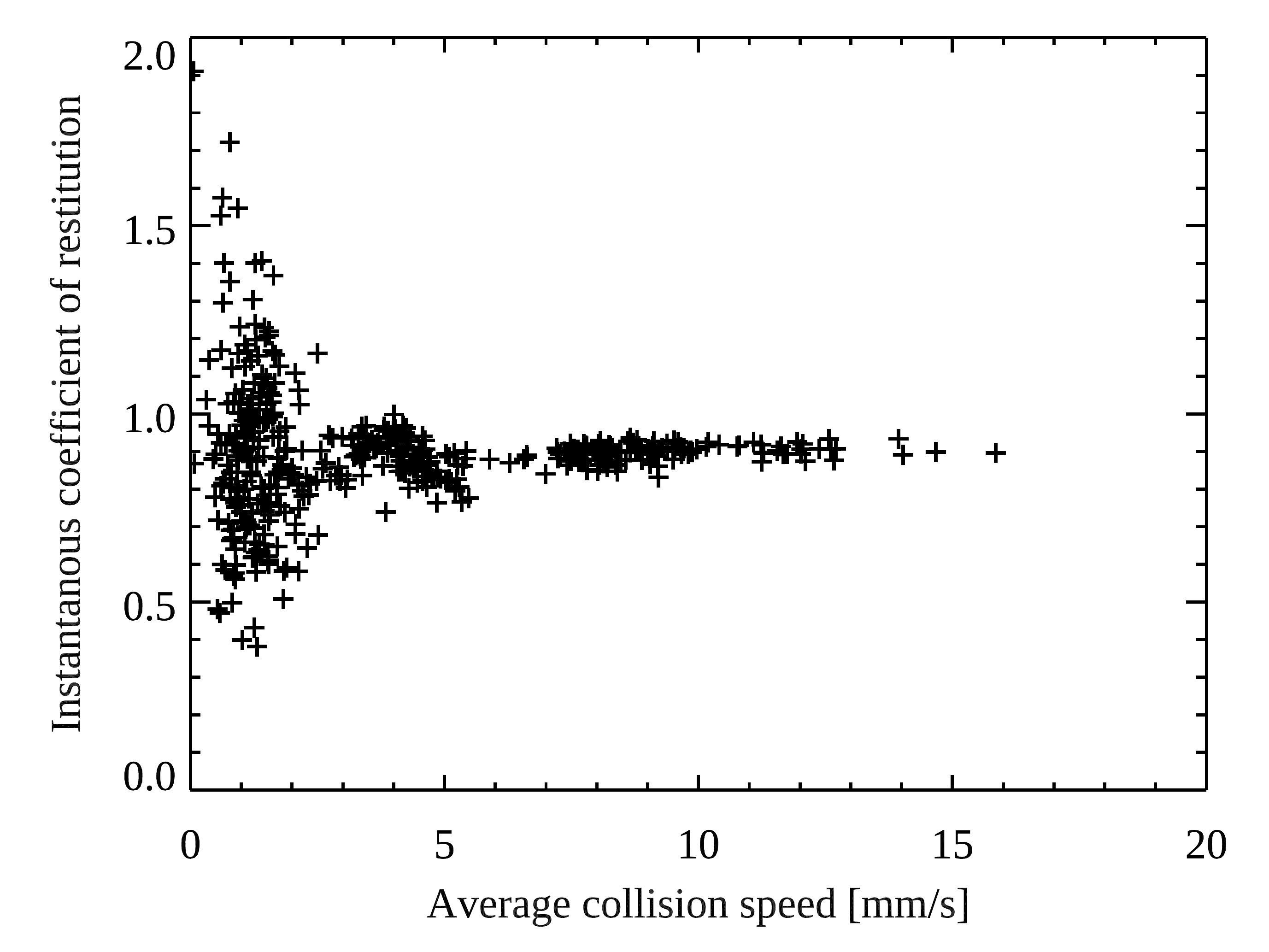}
 \caption{Instantaneous coefficient of restitution in tray~1 computed from the averaged velocity profiles of all NanoRocks experiment runs. The coefficient of restitution average remains around 0.89 independently from the current average collision speed, with a narrow distribution around the mean for collision speeds above 5 mm/s. For collision speeds at or below 5 mm/s, the width of the distribution of velocities increases significantly.}
 \label{f:cor_speed}
 \end{center}
\end{figure}

Having obtained the instantaneous coefficient of restitution, we can correlate it with the instantaneous collision speed (twice the average particle speed) in the experiment cell. In Figure~\ref{f:cor_speed}, we show the results of this analysis for the NanoRocks tray~1. We observe that the instantaneous coefficient of restitution is constant in average and not velocity dependent in our experiment. For collision speeds above about 5 mm/s, the variations around the mean value are low. Below 5 mm/s, the variations around the mean are larger and increase with decreasing collision speed. As we averaged our measurements over 30 frames, values under 0.5 mm/s can show inaccuracies. However, the trend to a wider distribution of coefficients of restitution can still be identified. Coefficients of restitution larger than 1 are commonly obtained when particles are rotating before the collision, in which case some of the rotational energy is converted into a translational component upon the collision.

From this data, we can confirm that our assumption of a velocity-independent coefficient of restitution in our statistical analysis is valid (constant average in Figure~\ref{f:cor_speed}). In addition, we observe that the exchange of energy between translational and rotational motion becomes efficient at relative velocities below 2~mm/s, with maximum coefficients of restitution of 2. This indicates that both translational and rotational energies are of the same order. This allows to estimate the average rotational velocity of the grains to be around 0.1$\degree$/s in this phase of the experiment. 

Compared to the data collected in \cite{heisselmann2010microgravity}, the distribution of values observed is much narrower for speeds above 5 mm/s. This difference could originate in the statistical nature of our data analysis, while coefficients of restitution were measured for individual collisions in \cite{heisselmann2010microgravity}. 

\subsection{Collisions and Growth of Dust in Protoplanetary Disks}

\paragraph{\textit{Hierachical Growth of Dust Particles}}

The study of low-velocity collisions between mm-sized particles is of particular relevance to the growth of dust grains in PPDs. As shown by previous experiments \citep{guettler_et_al2010A&A,weidling_et_al2012Icarus,kothe_et_al2013Icarus} and simulations \citep{zsom_et_al2010AA}, hierarchical growth of dust particles is stalled by the "bouncing barrier" once the particles reach mm to cm sizes. Thorough studies of the collision behavior of mm to cm-sized particles in the low-velocity regimes encountered in the PPD ($<$ 1-10 cm/s) thus supports our understanding of grain size evolution in PPDs and eventually the formation of larger planetesimals.

\begin{figure}[t]
  \begin{center}
  \includegraphics[trim=20 20 15 0, clip, width = 0.48\textwidth]{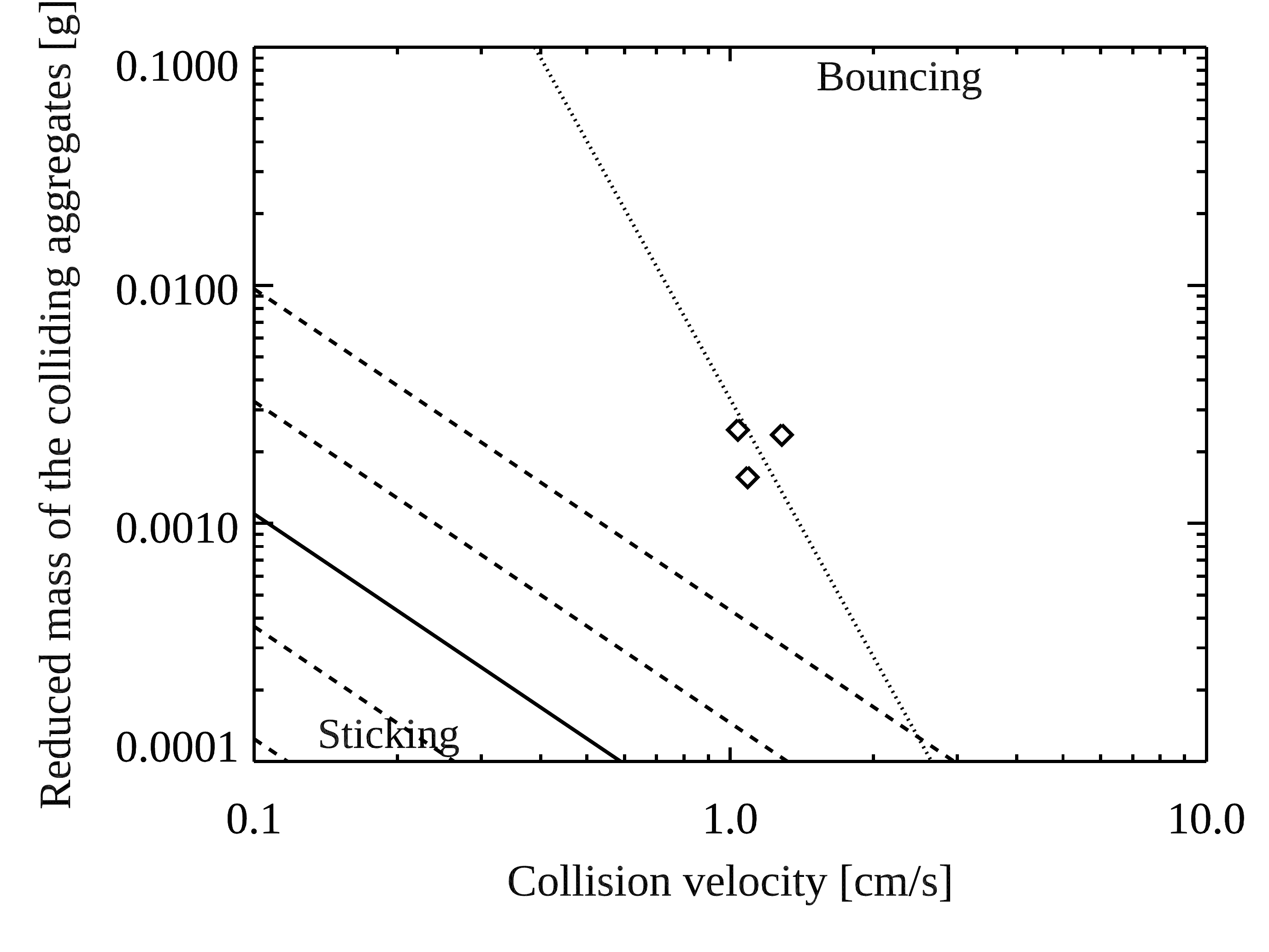}
 \caption{NanoRocks sticking threshold velocity compared to the dust collision models by \cite{weidling_et_al2012Icarus} and \cite{kothe_et_al2013Icarus}. The three points shown (diamonds) are for particles in trays 1, 3, and 5. The dotted line is the 100 \% bouncing limit from \cite{weidling_et_al2012Icarus}. The solid and dashed lines are the 100, 75, 50 (solid), 25, and 0 \% bouncing limits from top to bottom.}
 \label{f:model}
 \end{center}
\end{figure}

In Figure~\ref{f:model}, we compare the sticking threshold velocities measured in the NanoRocks experiment to current dust collision models by \cite{weidling_et_al2012Icarus} (dotted line) and \cite{kothe_et_al2013Icarus} (dashed and solid lines). The diamonds represent the data points from trays 1, 3, and 5, for which reliable clustering onset data could be measured (see \ref{s:onset}). As mentioned in \ref{s:onset}, the collision speed (twice the average particle speed) was measured at the onset of clustering, which is linked to the sticking threshold for inter-particle collisions. Values obtained were 1.04, 1.09 and 1.28~cm/s for particles in trays 1, 3 and~5, respectively. These data are clustering around the 100 \% bouncing limit computed by \cite{weidling_et_al2012Icarus} (dotted line), indicating that our statistical method for determining the onset of clustering (described in \ref{s:onset}) detects the very first sticking events and is therefore detecting the proper \textit{onset} of sticking rather than the average sticking threshold velocity. The good agreement with the \cite{weidling_et_al2012Icarus} model also indicates that the NanoRocks particles, which have higher masses than in \cite{weidling_et_al2012Icarus} but are similar in their structure (individual particles rather than aggregates of particles), behaved as predicted by this model. The \cite{kothe_et_al2013Icarus} model was computed using measurements on collisions between aggregates of particles, some of which had elongated and fractal shapes. During the homogeneous cooling phase of the NanoRocks experiment runs, we only expect particle-particle collisions, as aggregates are not forming in the bouncing regime. Our statistical method for measuring the first sticking events did not allow for the detection of particle-aggregate collisions, so that the onset of clustering speeds measured in NanoRocks are not expected to match the \cite{kothe_et_al2013Icarus} model. 

The good agreement of the NanoRocks data with the \cite{weidling_et_al2012Icarus} model also indicates that our experiment setup and the statistical data analysis presented here reliably reproduce data collected manually with a different experiment setup, thus validating our data analysis method and the resulting measurements. It also supports the dust collision models predicting the bouncing regime for mm-sized particles and collisions at speeds above a cm/s. This means that hierarchical particle growth in the PPD can only proceed for mm-sized monomers in extremely quite environments where relative velocities are damped below a few cm/s. While such low speeds are easily obtained for very small ($\mu$m) particles due to their nearly perfect coupling to the ambient gas, larger grains tend to decouple from the gas, increasing relative velocities. In addition, various gas turbulences (e.g. Magneto-Rotational Instability, \cite{balbus1991powerful}) and instabilities (e.g. Kelvin-Helmholtz instability, \cite{youdin2002planetesimal,lee2010forming}) in the PPD keep relative velocities from settling to low values. Therefore, growth beyond the mm size scale would have to proceed at higher collision speeds, which might require either very fractal and porous aggregates \citep{okuzumi2012rapid} or large size ratios between colliding partners \citep{windmark_et_al2012AA}. As the NanoRocks setup was not well-suited for reproducing pristine aggregate growth (immediate compaction due to collisions with the cell walls, see \ref{s:morph}), the study of particle growth beyond the mm scale would require different experimental setups. 

\paragraph{\textit{Particle Cloud Collapse}}

One of the mechanisms currently considered viable for the formation of $>$ km-sized planetesimals despite the various growth barriers is the concentration and gravitational collapse of dust particle clouds in the disk. Such concentrations can be due to particle trapping in turbulent Eddies \citep{barge1995did,meheut2012dust} or streaming instabilities \citep{youdin_goodman2005,Johansen_et_al2007ApJ,bai2010dynamics}. Numerical simulations of such collapses show that the bouncing regime plays a significant role in the collapse time scales, therefore constraining the disk and particle parameters for which planetesimals can be formed consistently. \cite{jansson2014formation} show that for low-mass clouds (eventually forming bodies of $\sim$ 5 km in size), the potential energy of the falling particles is efficiently dissipated by bouncing collisions. In this case, the collapse time is proportional to $(1-\epsilon^2)^{-1}$, with $\epsilon$ being the inter-particle coefficient of restitution. For simplicity, \cite{jansson2014formation} assumed perfectly inelastic collisions ($\epsilon = 0$). The consideration of a non-zero coefficient of restitution, as observed in the NanoRocks and other experiments \citep[e.g.][]{blum_and_muench1993Icarus,weidling_et_al2012Icarus,weidling2015free}, would increase the cloud collapsing times by up to a factor 5 (if we take $\epsilon = 0.89$ as measured in the NanoRocks tray~1). This in turn influences the very possibility of planetesimal formation, as the collapse time is in competition with the timescale for inward drifting of the particles due to gas drag in the PPD \citep{weidenschilling1977MNRAS}. 

In the case of larger clouds (final body sizes above 70~km), \cite{jansson2014formation} find that the increased collision speeds inside the cloud lead to fragmentation, thus accelerating the collapse as the particle number density and collision rates increase. In this case, fragmentation products are generated in the form of a population of fine dust grains (0.1 mm). In the inset of Figure~\ref{f:kin_en}, we see that the presence of fine grains in the particle population of tray~2 significantly reduces the measured coefficient of restitution, which would lead to a shorter collapse time of the cloud. Even though restitution coefficients were ignored in \cite{jansson2014formation}, their consideration would also support the trend to shorter collapsing times in the case of fragmenting collisions in the cloud.

\subsection{Particle collisions in planetary rings}

The collision velocities in NanoRocks are comparable to those in unperturbed regions of dense planetary rings, such as the main rings of Saturn and the narrow dense rings of Uranus. However, the particle sizes in NanoRocks are smaller than the particle size distributions in those rings which typically range from about a centimeter up to several meters. Nevertheless, our conclusion that using a single average coefficient of restitution to represent a stochastic distribution of coefficients of restitution is applicable to N-body simulations of planetary rings. Our results suggest that single-valued coefficients of restitution are appropriate for such simulations and that velocity-dependent models of the coefficient of restitution should not be uniformly applied in ring simulations. If the velocity-dependent effect observed in previous experiments is in fact due to an effectively infinite mass of one of the colliding objects, then simulations may need to incorporate velocity-dependent coefficients of restitution for collisions between particles of very different masses, but a single value for collisions between like-sized particles.

\subsection{Limitations of the NanoRocks experiment setup}

In the discussion above, we have shown that, with the exception of trays containing JSC-1 grains, our experiment setup was able to reproduce homogeneously cooling systems of multiple particles fairly well. The measurement of the onset of particle clustering was also in good agreement with predictions of current dust collision models. In the case of 100~$\mu$m-sized grains, the shaking mechanism was not powerful enough to de-aggregate large clusters efficiently.

In trays where homogeneously cooling particle systems were well reproduced, a major limitation was also the shape and structure (porosity) of the clusters formed during the experiment runs. These were strongly influenced by the hardware design: packing fractions were at the limits of the highest possible values (see Figure~\ref{f:ff}) and clusters were immediately compacted due to collisions with the cell walls. NanoRocks was therefore well suited for the study of the bouncing regime of mm-sized grains, the damping of kinetic energy through multiple collisions, and the timing of clustering collapse of the system, but it is not possible to grow pristine aggregates and study their original structure. Therefore, NanoRocks, like any other experiment constraining and agitating particles via cell walls, is limited in its possibilities to study aggregate growth. Experimental setups using non-contact technologies for constraining and agitating particles will be better suited for such future studies.

\section{Summary}

In this paper, we present the results of the NanoRocks experiment, which studied the collisional evolution of multi-particle systems on-board the ISS for 18 months. This miniaturized experiment recorded collisions between 100 $\mu$m to 2 mm particles in 8 different experiment cells after regular agitation events. We studied the damping of kinetic energy through collisions and the formation of clusters. Due to the large number of collisions recorded, we developed statistical analysis methods in order to collect quantitative information from the recorded data. This allowed us to measure collision restitution coefficients, sticking threshold velocities and cluster filling factors in several of the experiment cells. In order to validate these measurements, we performed numerical simulations reproducing the NanoRocks particle systems and compared the results to our experimental data. We obtained the following findings, which have the potential to support future numerical simulations of collisional multi-particle systems:

\begin{itemize}

	\item The average coefficient of restitution of collisions between same-sized free-floating particles at low speeds ($<$ 2 cm/s) is not dependent on the collision velocity. This is in contradiction to former experimental and numerical studies, but can be reconciled by considering the details of the experimental setups: collisions of particles with flat surfaces have velocity-dependent restitution coefficients, while collisions between free-floating particles have constant restitution coefficients;
	\item The evolution of the particle average speeds during the collisional cooling phase is insensitive to the width of the distribution of coefficients of restitution. This means that the time evolution of the system is the same wheather a constant or a uniform distribution of values is used for the inter-particle restitution coefficients. The simplified approach of using a constant value for a specific type of particles is therefore valid and will accurately reproduce the average behavior of a multi-particle system during collisional cooling;
	\item The width of the distribution of coefficients of restitution around the average value increases with decreasing collision speed. At speeds below $\sim$ 5 mm/s, we observed that the coefficients of restitution between free-floating mm-sized particles is randomized around their average value, including values above unity. We conjecture that very gentle collisions allow for the transfer between rotational and translational kinetic energy, an effect that is supported by microscopic asperities at the surfaces of particles, thus producing a much wider range of values for the measured (translational) coefficient of restitution;
	\item In very quiescent environment (average collision velocities around 1~mm/s), the kinetic energy is stored in equipartition between translational and rotational motion, while the influence of particle rotation can be neglected for average speeds above 1 cm/s;
	\item Our measurements of the sticking threshold speeds for mm-sized particles are in good agreement with the current dust collision model based on data from different experimental setups. We conclude that such models are valid for the investigated particle size and mass ranges, thus confirming that the bouncing regime is dominant at collision speeds expected in the PPD for these particle sizes (chondrule type of particles), stalling hierarchical grain growth.

\end{itemize}

In addition, we pointed out the limitations of the NanoRocks experiment setup. The analysis of the packing density of the clusters formed in our experiment cells showed rapid compaction by collisions with the experiment cell walls, destroying the original structure of the clusters. Therefore, NanoRocks is best suited for the study of the bouncing regime and its limits, rather than the study of the formed clusters.

\begin{acknowledgements}
This work is based in part upon research supported by NASA through the Origins of Solar Systems Program grant NNX09AB-85G and by NSF through grant AST-1413332. The NanoRocks experiment was supported by Space Florida, NanoRacks LLC, the Florida Space Institute, and the University of Central Florida.
\end{acknowledgements}

\end{document}